\documentclass[a4paper,preprint, pre,showpacs]{revtex4}

\usepackage{amsmath,amssymb}
\usepackage{makeidx}
\usepackage{amsfonts}
\usepackage[ansinew]{inputenc}
\usepackage[usenames,dvipsnames]{pstricks}
\usepackage{epsfig,graphicx}

\DeclareMathOperator*{\sgn}{sgn}

\newcommand{\minus}{{\!-\!}}	
\newcommand{\vecSpin}{\mbox{\boldmath$S$}}
\newcommand{\vecSpinhat}{\hat\vecSpin}
\newcommand{\vecPsi}{\mbox{\boldmath$\psi$}}
\newcommand{\vecPsihat}{\hat\vecPsi}

\newcommand{\Spinhat}{\hat{\Spin}}
\newcommand{\Spin}{\mbox{$S$}}
\newcommand{\vecProb}{\mbox{\boldmath$P$}}
\newcommand{\vecProbhat}{\hat\vecProb}
\newcommand{\Prob}{\mbox{$P$}}
\newcommand{\Probhat}{\hat{\Prob}}
\newcommand{\M}{\mbox{$M$}}
\newcommand{\Omegahat}{\hat\Omega}
\newcommand{\vecOmega}{\mbox{\boldmath $\Omega$}}
\newcommand{\vecOmegahat}{\hat\vecOmega}
\newcommand{\vecField}{\mbox{\boldmath $H$}}
\newcommand{\vecconjField}{\mbox{\boldmath $x$}}
\newcommand{\vecomega}{\mbox{\boldmath $\omega$}}
\newcommand{\conjField}{\mbox{$x$}}
\newcommand{\Field}{\mbox{$H$}}
\newcommand{\Fieldhat}{\hat{\mbox{$H$}}}
\newcommand{\conjFieldhat}{\hat{\conjField}}
\newcommand{\Psihat}{\hat\psi}
\newcommand{\betahat}{\hat\beta}
\newcommand{\nullvec}{\mbox{\boldmath$0$}}
\newcommand{\Z}{\mathbb{Z}}
\newcommand{\rme}{\mathrm{e}}
\newcommand{\rmi}{\mathrm{i}}
\newcommand{\rmd}{\mathrm{d}}

\begin{document}

\title{Noisy Random Boolean Formulae - a Statistical Physics Perspective}

\author{Alexander Mozeika$^1$}
\author{David Saad$^1$}
\author{Jack Raymond$^2$}

\affiliation{ $^1$The Non-linearity and Complexity  Research Group, Aston University, Birmingham B4 7ET, UK.\\ $^2$Department of Physics, The Hong Kong University of Science and Technology, Clear Water Bay, Hong Kong, China.}
\email{a.s.mozeika@aston.ac.uk}

\date{\today}

\begin{abstract}
Typical properties of computing circuits composed of noisy logical gates are studied using the statistical physics methodology. A growth model that gives rise to typical random Boolean functions is mapped onto a layered Ising spin system, which facilitates the study of their ability to represent arbitrary formulae with a given level of error, the tolerable level of gate-noise, and its dependence on the formulae depth and complexity, the gates used and properties of the function inputs. Bounds on their performance, derived in the information theory literature via specific gates, are straightforwardly retrieved, generalized and identified as the corresponding typical-case phase transitions. The framework is employed for deriving results on error-rates, function-depth and sensitivity, and their dependence on the gate-type and noise model used that are difficult to obtain via the traditional methods used in this field.
\end{abstract}



\maketitle

\section{Introduction\label{section:Intro}}

Computation as a physical phenomena takes many forms including classical logical circuits, quantum computing and biological neural networks. Noise is present in all practical computing systems and is a source of error with an immediate effect on the ability to represent specific functions and operations and the viability of some computing paradigms. The main sources of errors in classical computing circuits based on semiconductor technology are heat generation, cosmic rays and production defects~\cite{Borkar}. The impact of noise becomes even more dramatic as the drive towards miniaturization of computer components causes the circuits to become more complex and of large scale~\cite{Borkar}. The presence of decoherence-noise in quantum computers is also a significant obstacle for exploiting their full computational power~\cite{QuantumCompReview}. The effects of noise on computing and information processing in other systems, such as biological neural networks, which are inherently noisy, remain poorly understood.

One of the first to study the effect of noise in computing systems was von Neumann~\cite{VonNeumann} who attempted to explain the robustness of biological computing circuits by representing them as logical circuits comprising conventional Boolean logical gates. His model represented neural activities by a circuit (or formula) composed of $\epsilon$-noisy Boolean gates; he suggested alternative gate-constructions to limit the resulting noise and analyzed the maximal noise tolerated before the generated functions become random. 

Before progressing any further, a few formal definitions are required: (i) A \emph{circuit} may be regarded as a directed acyclic graph in which the nodes of in-degree zero are either Boolean constants or references to arguments, the nodes of in-degree $k\geq1$ are logical gates of $k$ arguments and the nodes of out-degree zero correspond to the circuit outputs. (ii) A \emph{formula} is  a single-output circuit where the output of each gate is used as an input to at most one
gate. (iii) The  $\epsilon$-noisy gate is designed to compute a Boolean function $\alpha:\{-1,1\}^k\rightarrow\{-1,1\}$, but for each input $\vecSpin\in\{-1,1\}^k$  there is an error probability $\epsilon$ such that $\alpha(\vecSpin)\rightarrow-\alpha(\vecSpin)$. To simplify the analysis, error-probability is taken to be independent for each gate in the circuit. Clearly, a noisy circuit ($\epsilon>0$) cannot perform any given computation in a deterministic manner: for any circuit-input there is a non-vanishing probability that the circuit will produces the wrong output. (iv) The maximum of this error probability $\delta$ over all circuit-inputs determines \emph{reliability} of the circuit. 

In his paper, von Neumann showed that reliable computation ($\delta<1/2$) is possible for a sufficiently small $\epsilon$~\cite{VonNeumann} and demonstrated how reliability of a Boolean noisy circuit can be improved by using gate-constructions based only on $\epsilon$-noisy gates. There had been little development in the analysis of noisy computing systems until the seminal work of Pippenger~\cite{Pippenger:RC} who addressed the problem from an information theory point of view. He showed that if a noisy $k$-ary formula is used to compute a Boolean function $f$ with the error probability $\delta<1/2$, then (i) there is an upper bound for the gate-error $\epsilon(k)$ which is strictly less than $1/2$ and (ii)
there is a lower bound for the formula-depth $\hat{d}(k,\epsilon,\delta)\ge d$, where $d$ is the depth of a noiseless formula computing $f$; the \emph{depth} of a formula being the number of gates on the longest path from an input node to the output node. In comparison to its noiseless counterpart, a noisy formula that computes reliably has greater depth due to the presence of restitution-gates, implying longer computation times~\cite{Pippenger:RC}. 

A number of papers have followed and extended Pippenger's results. For instance, similar results were derived for circuits by Feder~\cite{Feder:RC}, who also improved the bounds obtained by Pippenger for formulae. The exact noise thresholds for $k$-ary Boolean formulae were later determined in~\cite{Hajek:MTN},~\cite{Evans:MTNK} (for odd $k$ only). The source of the parity restriction on $k$ originates from use of a specific gate in the corresponding proofs, the majority gate (MAJ-$k$), for constructing noisy formulae; these gates have shown to be optimal for preserving a single input-bit of information~\cite{Evans:MTNK}. For formulae constructed from gates with an even number of inputs only the noise threshold for $2$-input NAND gate formulae was computed exactly~\cite{Evans:MTN}. A recent result~\cite{Unger} suggests that this threshold is an exact noise threshold of general $2$-input gate formulae. 

Against this plethora of results from the information theory and theoretical computer science literature, our aim is to provide an alternative view based on a statistical physics framework, which we believe offers a powerful methodology that can recover and extend existing results to provide insight beyond what is accessible via the information theory methodology. 
The latter mainly rely on specific circuit constructions and methods that correspond to the \emph{worst} case bounds. 
In contrast, our emphasis is on the \emph{typical} case analysis of noisy circuits, which facilitates the study of properties at any depth and offers flexibility in extending the results to any distribution of logical gates.

The analysis in the current paper is based on path integral methodology, specifically tailored for this task and originated in the statistical physics of disordered systems. It complements other methods that have been successfully employed in the study of similar problems from theoretical computer science and information theory~\cite{IPCbook} ranging from classical combinatorial optimization problems (graph coloring, k-SAT, reconstruction on trees and graph-isomorphism to name but a few) to source and channel coding~\cite{ldpcc}, but is arguably more appropriate here due to the directed nature of the formulae studied. As in the previous cases, we believe that our understanding will be significantly enhanced by interaction across disciplines~\cite{IPCbook}. 

The study of noisy computing requires the generation of {\em typical} functions. Apparently, constructing typical functions by randomly connecting Boolean gates is not trivial and constitutes an area of research on its own right. Most of the familiar paradigms in the theoretical computer science literature identify gates or processes that can represent any arbitrary function, but when applied at random they tend  generate trivial functions showing weak dependence on the input variables.
To generate typical formulae, which compute all Boolean functions with uniform probability, using \emph{randomly generated} circuits, we employed a variant of the growth process suggested by Savick\'{y}~\cite{Savicky} that, under very broad conditions, produces typical functions as the depth of the formulae becomes large.

The remainder of this paper is organized as follows. In section~\ref{section:RBF} we discuss generation of typical Boolean functions and the model we employ for generating them. In section~\ref{section:Model} we define our model of noisy computation used for the analysis followed by the derivation of the corresponding mean-field theory in section~\ref{section:Method}. Results obtained by applying our method to random formulae which use single gates or distribution over gates are presented in section \ref{section:Results} followed by a summary and discussion of future work in section~\ref{section:Discussion}. Technical aspects of the calculations which lead to our theoretic results are provided in the Appendixes~\ref{appendix:Process}-\ref{appendix:SPproblem}.

\section{Random Boolean Functions\label{section:RBF}}

To investigate the effect of gate-noise on circuits representing random Boolean functions one should first identify a method for generating such circuits using basic logical gates. The importance of random Boolean functions is in the fact that they  facilitate the study of average case properties, in contrast to the traditionally-studied worst-case scenario~\cite{Brodsky}. 

A common approach to represent a random Boolean function is by constructing a random Boolean circuit or formula. However, finding
a circuit representation of a Boolean function using a particular set of gates and of a bounded size is considered a difficult problem. The majority of methods designated for this task use covering or bi-decomposition as their basic procedure~\cite{Steinbach}. Applying the covering method results in a disjunctive normal form (DNF) representation. The DNF or its dual CNF (conjunctive normal form) is a depth-2 formula with AND and OR gates used as internal nodes and with the input Boolean variables and their negations distributed on the leaves. However, random DNF (CNF) formulae offer very low sensitivity~\cite{Boppana} to the input values and any attempt to construct them at random is likely to produce a highly uncharacteristic random Boolean functions.

Another approach  is based on a sequential bi-decomposition of the random function to be implemented. In this approach, one finds a circuit representation of the Boolean function by reducing its dependence on a single variable at each branch of a tree, sequentially. At each step, the Boolean function is decomposed into two simpler functions, of the remaining variables, that consider the two possible values of the given variable; this procedure is repeated until a circuit representation is found. The resulting representation may be suboptimal and it is not clear how to randomize this procedure in order to produce typical Boolean functions for a given set of simple gates. 

The most studied methods of generating random Boolean functions use random tree generation or a growth process as their core procedure. We will briefly introduce two of these methods. In the first method, a rooted $k$-ary tree is sampled from the uniform distribution of all rooted $k$-ary trees; the leaves of this random tree are then labeled by the reference to the Boolean variables and the internal nodes are labeled by the Boolean gates used. Lefmann and Savick\'{y} used this construction to investigate typical properties of large random Boolean AND/OR formulae~\cite{Lefmann} and obtained bounds on the probability $\Prob(f)$ for a random formula to compute a given Boolean function $f$. These bounds  were improved in follow-up studies~\cite{Chauvin,Gardy}, which also showed that for a small number of inputs $n$ the AND/OR model results in very simple functions~\cite{Chauvin} with high probability. They also suggested that this behavior becomes even more pronounced for large $n$.

The second method uses the following growth process: Firstly, one defines an initial distribution over a set of simple Boolean functions. Secondly, and in further steps, the formulae chosen from the distributions defined in previous steps are combined by Boolean gates. One such process, described by Savick\'{y}~\cite{Savicky}, uses only a single Boolean gate $\alpha$ and is defined by the recursion on the set of formulae $A_\ell$:
\begin{eqnarray}
&&A_0=\{1,-1,\Spin_1,\ldots,\Spin_n,-\Spin_1,\ldots,-\Spin_n,\}\nonumber\\
&&A_{\ell+1}=\{\alpha(\phi_1,\ldots,\phi_k);\phi_j\in A_\ell \textrm{ for } j=1,2,\ldots,k\}.\label{def:growth-process}
\end{eqnarray}
Savick\'{y} showed, under a very broad conditions on $\alpha$, that the probability of computing a Boolean function by a formula $\phi\in A_\ell$ tends to the uniform distribution over all Boolean functions of $n$ variables when $\ell\rightarrow\infty$~\cite{Savicky}. Furthermore, depending on the initial conditions $A_0$ and the gate $\alpha$ the process converges to a {\em single Boolean function} or to the {\em uniform distribution} over some class of Boolean functions~\cite{Brodsky}.

\begin{figure}[t]
\vspace*{-11mm} \hspace*{-0mm} \setlength{\unitlength}{0.25mm}
\begin{picture}(350,210)
\put(0,0){\includegraphics[height=160\unitlength,width=280\unitlength]{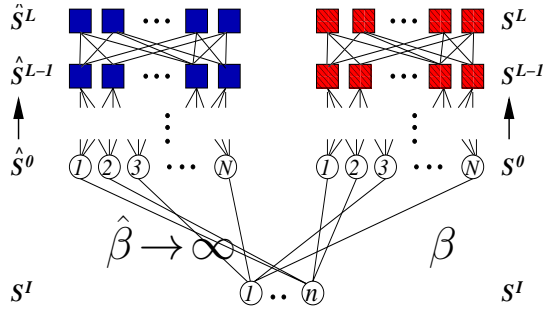}}
\put(50,25){\Large{$\hat{\beta}\!\rightarrow\!\infty$}}
\put(220,25){\Large{$\beta$}}
\end{picture}
 \vspace*{0mm}
\caption{(Color online) The model of two coupled systems with identical topology and different inverse temperatures $\beta$ and $\hat{\beta}\!\rightarrow\!\infty$. Gates are indicated by squares, $S^I$ and input nodes by circles. Blue indicates noiseless gates, red noisy gates. \label{fig:0} \vspace*{-0.3cm}
}
\end{figure}

In this framework, all Boolean functions of $n$ variables are represented with equal statistical weight when $\ell\rightarrow\infty$, but the number of gates in formulae grows exponentially with the formula depth $\ell$. Here, in order to tame this explosion in the number of gates, we propose a layered variant of the Savick\'{y} growth process. The first step in our process is to sample randomly and uniformly exactly $N$ entries of an input vector $\vecSpinhat^0=(\Spin_1^0,\ldots,\Spin_N^0)$. In the second, and all subsequent steps for $\ell=1,\ldots,L-1$, we construct a vector $\vecSpinhat^{\ell+1}=(\Spin_1^{\ell+1},\ldots,\Spin_N^{\ell+1})$ where the $i$-th entry $\Spin_i^{\ell+1}$ is an output of the gate $\alpha(\Spin_{i_1}^{\ell},\ldots,\Spin_{i_k}^{\ell})$ with $k$ input-indices sampled uniformly from the set of all possible (unordered) indices $\{i_1,\ldots,i_k\}$. The result of the process is the layered $N\times(L+1)$ Boolean circuit shown in Figure~\ref{fig:0} (left construction, in blue). For large $N$, the variable $\Spin_i^{\ell}$ in our model corresponds to the output of a random $k$-ary of depth $\ell$, which computes a Boolean function $\{-1,1\}^{N}\rightarrow\{-1,1\}$. We expect that in the limit $N\rightarrow\infty$, with $\ell\in O(N^0)$,  the statistical properties of the formulae generated in our process  and in the Savick\'{y}'s growth process are equivalent; this is supported by the results reported later. The advantage of using the layered representation is that it allows us to explore the typical behavior of noisy random Boolean formulae using methods of statistical physics. 

While the vector $\vecSpinhat^0$ represents randomly sampled single entries, one would also like to study cases where entries are statistically dependent and are sampled from a smaller set. To cater for a possible higher level of correlation, the $0$-layer boundary conditions are generated by selecting randomly $\vecSpinhat^0$ entries from members of the finite set  $S^I=\{S_1^I,\ldots,S^I_n\}$. This allows to investigate the properties of the functions generated and their dependence on properties of the set $S^I$.

\section{Model\label{section:Model}}

As described in section \ref{section:RBF}, the noisy computation model consider here is a feed-forward layered $N\times (L+1)$ Boolean circuit. The layers in the circuit are numbered from $0$ (input) to $L$ (output). Each layer $\ell\in\{1,\ldots,L\}$ in the circuit is composed of exactly $N$ $\epsilon$-noisy, $k$-ary Boolean gates. Noise at gate $\alpha_i^\ell$ on site $(i,\ell)$ operates independently and in a stochastic manner according to the microscopic law (see Appendix \ref{appendix:Process})
\begin{eqnarray}
\Prob(S_{i}^\ell\vert S_{i_1}^{\ell-1},\ldots,S_{i_k}^{\ell-1})&=&\frac{\rme^{\beta S_{i}^\ell\alpha_i^\ell(S_{i_1}^{\ell-1},\ldots,S_{i_k}^{\ell-1})}}{2\cosh[\beta\alpha_i^\ell(S_{i_1}^{\ell-1},\ldots,S_{i_k}^{\ell-1})]}~,\label{eq:micro}
\end{eqnarray}
where $\beta=1/T$ is the inverse temperature, related to the gate noise $\epsilon$ via $\tanh\beta=1-2\epsilon$. The gate-output $\Spin_{i}^\ell$ is completely random when $\beta\rightarrow0$ ($\epsilon=1/2$) and completely deterministic when $\beta\rightarrow\infty$ ($\epsilon=0$). Our model is acyclic by definition, so given the state of gates at layer $\ell$, gates at layer $\ell+1$ operate independently of each other. The latter suggests that the probability of the microscopic state $\vecSpin^0,\ldots,\vecSpin^L$, where $\vecSpin^\ell\in\{-1,1\}^N$, is just a product of equation~(\ref{eq:micro}) over all sites and layers in the circuit. Furthermore, to investigate the properties of noisy formulas we consider two copies of the same topology, shown in Figure \ref{fig:0}, but with different temperatures $\beta<\infty$ (noisy) and $\hat\beta\rightarrow\infty$ (noiseless), comparing the two will enable us to study the effect of noise on the resulting functions. Following similar arguments to those of the single circuit case, the probability of microscopic states in the two systems are given by  
%
%
\begin{eqnarray}
\Prob[\{\vecSpin^\ell\};\{\vecSpinhat^\ell\}]&=&\Prob(\vecSpin^0,\vecSpinhat^0\vert \vecSpin^I)\prod_{\ell=1}^L \Prob(\vecSpin^\ell\vert\vecSpin^{\ell-1})P(\vecSpinhat^\ell\vert\vecSpinhat^{\ell-1})\label{eq:PathProb}
\end{eqnarray}
where 
\begin{eqnarray}
\Prob(\vecSpin^\ell\vert\vecSpin^{\ell-1})&=&\prod_{i=1}^N\frac{\rme^{\beta S_{i}^\ell\sum_{j_1,\ldots,j_k}^N A_{j_1,\ldots,j_k}^{\ell,i}\alpha_i^\ell(S_{j_1}^{\ell-1},\ldots,S_{j_k}^{\ell-1})}}{2\cosh[\beta\sum_{j_1,\ldots,j_k}^N A_{j_1,\ldots,j_k}^{\ell,i}\alpha_i^\ell(S_{j_1}^{\ell-1},\ldots,S_{j_k}^{\ell-1})]}.\label{eq:layerProb}~.
\end{eqnarray}
 %
 %
The set of connectivity tensors $\{A_{i_1,\ldots,i_k}^{\ell,i}\}$,
where $A_{i_1,\ldots,i_k}^{\ell,i}\in\{0,1\}$,  denotes connections in the circuit. The conditional probability $\Prob(\vecSpinhat^\ell\vert\vecSpinhat^{\ell-1})$ is the same as in equation~(\ref{eq:layerProb}) but with $\beta\rightarrow\betahat$.

 The sources of disorder in our model are the random connections, random boundary conditions and random gates. The former two arise in the layered growth process described in the last two paragraphs of section~\ref{section:RBF}. The basic step in this growth  process is the addition of a new gate with probability $\Prob(A_{j_1,\ldots,j_k}^{\ell,i})=\frac{1}{N^k}\delta_{A_{j_1,\ldots,j_k}^{\ell,i};1}+(1-\frac{1}{N^k})\delta_{A_{j_1,\ldots,j_k}^{\ell,i};0}$ of being connected to exactly $k$ gate-outputs on the previous layer $\ell-1$. This procedure is carried out independently for all gates in the circuit giving rise to the probability distribution 
\begin{eqnarray}
\label{def:connect-disorder-main}
\Prob(\{A_{i_1,\ldots,i_k}^{\ell,i}\})=\frac{1}{Z_A}\prod_{\ell,i=1}^{L,N}\left[\delta\left[1; \sum_{j_1,\ldots,j_k}^N A_{j_1,\ldots,j_k}^{\ell,i}\right]\prod_{i_1,\ldots,i_k}^{N}\left[\frac{1}{N^k}\delta_{A_{i_1,\ldots,i_k}^{\ell,i};1}+(1-\frac{1}{N^k})\delta_{A_{i_1,\ldots,i_k}^{\ell,i};0}\right]\right]\label{def:connect-disorder}
\end{eqnarray}
where $Z_A$ is a normalization constant. The Kronecker delta function inside the definition (\ref{def:connect-disorder-main}) enforces the constraint $\sum_{j_1,\ldots,j_k}^N A_{j_1,\ldots,j_k}^{\ell,i}=1$, i.e. the gate on site $(i,\ell)$ is mapped to exactly one element from the set of all possible output-indices $\{i_1,\ldots,i_k\}$ from the previous layer. Other sparse connectivity profiles can be easily accommodated into our framework by incorporating additional constraints into the definition (\ref{def:connect-disorder-main}) via the appropriate delta functions. 

Random boundary conditions in the layered growth process are generated by selecting members of the input set $S^I$, where $\vert S^I\vert\in O(N^0)$,  with probability $\frac{1}{\vert S^I\vert}$, and assigning them to the input layer $0$. The boundary condition is identical for two systems which leads to the distribution 
\begin{eqnarray}
\Prob(\vecSpin^0,\vecSpinhat^0\vert \vecSpin^I)=\prod_{i=1}^N \delta_{S^0_{i};S^I_{n_{i}}}\delta_{\hat{S}^0_{i};S^0_{i}}\label{def:initial}
\end{eqnarray}
where $\{n_i\}$ are independent random indices pointing to the members of input set $S^I$ with probability $P(n_{i})=\frac{1}{\vert S^I\vert}$. Further correlations can be introduced by defining the probability function $\Prob(S^I)$. 

In addition to the topological disorder, induced by the growth process, we assume that the gate $\alpha_{i}^{\ell}$ added at each step of the process can be sampled randomly and independently from the set $G$ of $k$-ary Boolean gates. Under this assumption the distribution over gates takes the form 
\begin{eqnarray}
\Prob(\{\alpha_{i}^{\ell}\})=\prod_{\ell,i=1}^{L,N}\Prob(\alpha_{i}^{\ell})\label{def:gate-disorder}
\end{eqnarray}
where $\Prob(\alpha_{i}^{\ell})=\sum_{\alpha\in G}p_{\alpha}\delta_{\alpha;\alpha_{i}^{\ell}}$ with $\sum_{\alpha\in G}p_{\alpha}=1$ and $p_{\alpha}\geq0$.

\section{Method\label{section:Method}}

To compute the probability distribution (\ref{eq:PathProb})
directly for a circuit of finite but significant size is
difficult. However, the structure of equation~(\ref{eq:PathProb})
is similar to the one that describes evolution of the disordered
Ising spin system~\cite{ParDyn}. This similarity becomes apparent
if one regards the layers in our model as discrete time-steps of
parallel dynamics. A common way to deal with the probabilistic
objects that take this form is to use the generating functional
method of statistical mechanics~\cite{dD}. The generating
functional for the current model
 is given by
\begin{eqnarray}
\Gamma[\vecPsi;\vecPsihat]&=&\left\langle\rme^{-\rmi\sum_{\ell,i}\{\psi_i^{\ell} S_{i}^{\ell}+\Psihat_i^{\ell} \hat{S}_{i}^{\ell}\}}\right\rangle\label{eq:GF}
\end{eqnarray}
where the shorthand $\langle\ldots\rangle$ denotes the average
over the joint probability~(\ref{eq:PathProb}). The generating
functional~(\ref{eq:GF}) can be regarded as a characteristic
function of~(\ref{eq:PathProb}) from which moments of the
distribution can be obtained by taking partial derivatives with
respect to the generating fields
$\{\psi_i^{\ell},\Psihat_j^{\ell^\prime}\}$, for example $\langle
S_i^{\ell}\hat{S}_j^{\ell^\prime}\rangle=-\lim_{\vecPsi,\vecPsihat\rightarrow\nullvec}
\frac{\partial^2}{\partial_{\psi_i^\ell}\partial_{\hat{\psi}_j^{\ell^\prime}}}\Gamma[\vecPsi;\vecPsihat]$.
Following prescription of~\cite{dD}, we assume that for
$N\rightarrow\infty$ the system is self-averaging and compute
$\overline{\Gamma[\vecPsi;\vecPsihat]}$, where
$\overline{[\cdots]}$ denotes an average over the disorder. The
disorder-averaged generating function~(\ref{eq:GF}) gives rise to
the following macroscopic observables
\begin{eqnarray}
&&m(\ell)\!=\!\frac{1}{N}\sum_{i=1}^N\overline{\langle S_i^\ell\rangle}=\!\!\lim_{\vecPsi,\vecPsihat\rightarrow\nullvec}\frac{\rmi}{N}\sum_{i=1}^N \frac{\partial}{\partial_{\psi_i^\ell}}\overline{\Gamma[\vecPsi;\vecPsihat]}\label{def:observ}\\ 
&&C(\ell)=\frac{1}{N}\sum_{i=1}^N\overline{\langle S_i^\ell\hat{S}_i^{\ell}\rangle}=-\lim_{\vecPsi,\vecPsihat\rightarrow\nullvec}\frac{1}{N}\sum_{i=1}^N \frac{\partial^2}{\partial_{\psi_i^\ell}\partial_{\Psihat_i^{\ell}}}\overline{\Gamma[\vecPsi;\vecPsihat]}\nonumber
\end{eqnarray}
 where $m(\ell)$ is the average activity (magnetization) on layer $\ell$ and $C(\ell)$ is the overlap between two systems. Averaging over the disorder in~(\ref{eq:GF}) leads to the saddle-point integral (see Appendix~\ref{appendix:Averages} for details)
\begin{eqnarray}
\overline{\Gamma}&=&\int\{\rmd \vecProb ~\rmd\vecProbhat ~\rmd
\vecOmega~ \rmd\vecOmegahat\}\rme^{N\Psi[\vecProb,\vecProbhat;
\vecOmega ,\vecOmegahat]}\label{eq:integral}
\end{eqnarray}
where $\Psi$ is the macroscopic saddle-point surface
%
%
\begin{eqnarray}
\Psi[\ldots]&=&\rmi\sum_{\ell=0}^{L-1}\sum_{ S,\hat{S}}\Probhat^\ell( S,\hat{S})\Prob^\ell( S,\hat{S})+\rmi\sum_{\ell=0}^{L-1}\int\rmd x ~\rmd \hat x~ \rmd\omega~\Omegahat^\ell(x,\hat x,\omega)\Omega^\ell(x, \hat x,\omega)\label{eq:saddle}\\
&+&\sum_{\ell=0}^{L-1}\sum_{\{S_j,\hat{S}_j\}}\prod_{j=1}^k\left[\Prob^\ell( S_j,\hat{S}_j)\right]\int\rmd x ~\rmd \hat x ~\rmd\omega~\Omega^\ell(x, \hat x,\omega)\left\langle\rme^{-\rmi\{x\alpha(\{S_j\})+\hat x\alpha(\{\hat{S}_j\})+\omega\}}\right\rangle_{\alpha}\nonumber\\
&+&\sum_n\Prob(n)\log\int\{\rmd\vecField~\rmd\vecconjField~\rmd\hat\vecField~\rmd\hat\vecconjField
\}\int\mathrm
D\vecomega\sum_{\vecSpin,\vecSpinhat}\M_n[\vecField,\vecconjField;\hat\vecField,\hat\vecconjField;\vecomega;\vecSpin,\vecSpinhat]~,
\nonumber
\end{eqnarray}
where $\langle\cdot\rangle_{\alpha}$ represents an average over
gate distribution and $\M$ is an effective single-site measure
\begin{eqnarray}
\M_n[\ldots]&=&\delta_{S^0;S^I_n}\delta_{\hat{S}^0;S^0}\prod_{\ell=0}^{L-1}\Big[\rme^{\rmi\conjField^\ell\Field^\ell+\rmi\conjFieldhat^\ell\Fieldhat^\ell +    \beta S^{\ell+1}\Field^{\ell}+\hat\beta\hat{S}^{\ell+1}\Fieldhat^{\ell}}\label{eq:M}\\
&&\times\rme^{-\log 2\cosh\left(\beta\Field^\ell\right)-\log
2\cosh\left(\hat\beta\Fieldhat^\ell\right)-\rmi\Probhat^\ell\left(
S^\ell,\hat{S}^\ell\right)-\rmi\Omegahat^\ell\left(\conjField^\ell,\conjFieldhat^\ell,\omega^{\ell+1}\right)+\rmi\omega^{\ell+1}}\Big].\nonumber
\end{eqnarray}
%
%
The generating fields $\vecPsi,\vecPsihat$ have been removed from
the above as they are no longer needed. For $N\rightarrow\infty$
the path-integral~(\ref{eq:integral}) is dominated by the extremum
of the functional $\Psi[\ldots]$ of equation~(\ref{eq:saddle}).
Functional variation of~(\ref{eq:saddle}) with respect to the
order parameters $\{\Prob,\hat{\Prob},\Omega,\hat{\Omega}\}$ gives
rise to four saddle-point equations
\begin{eqnarray}
&&\Prob^\ell( S,\hat{S})=\sum_n\Prob(n)\left\langle\delta_{ S^\ell; S}\delta_{\hat{S}^\ell;\hat{S}}\right\rangle_{M_n}\label{eq:SP1}\\
&&\Probhat^\ell( S,\hat{S})=\rmi\sum_{i=1}^k \sum_{\{S_j,\hat{S}_j\}}\delta_{ S_i; S}\delta_{\hat{S}_i;\hat{S}}\prod_{j\neq i}^k\left[\Prob( S_j,\hat{S}_j)\right]\label{eq:SP2}\\
&&~~~~~~~~~~~\times\int\rmd x ~\rmd \hat x~ \rmd\omega~\Omega^\ell(x, \hat x,\omega)\left\langle\rme^{-\rmi\{x\alpha(\{S_j\})+\hat x\alpha(\{\hat{S}_j\})+\omega\}}\right\rangle_{\alpha}\nonumber\\
&&\Omega^\ell(x,\hat x,\omega)=\sum_n\Prob(n)\left\langle\delta(\conjField-\conjField^\ell)\delta(\conjFieldhat-\conjFieldhat^\ell)\delta(\omega-\omega^{\ell+1})\right\rangle_{M_n}\label{eq:SP3}\\
&&\hat\Omega^\ell(x,\hat x,\omega)=\rmi\sum_{\{S_j,\hat{S}_j\}}\prod_{j=1}^k\left[\Prob^\ell( S_j,\hat{S}_j)\right]
\left\langle\rme^{-\rmi\{x\alpha(\{S_j\})+\hat x\alpha(\{\hat{S}_j\})+\omega\}}\right\rangle_{\alpha}\label{eq:SP4}
\end{eqnarray}
where $\langle\cdots\rangle_{M_n}$ is the average over the
probability distribution resulting from~(\ref{eq:M}). The saddle-point equations
(\ref{eq:SP1})-(\ref{eq:SP4}) can be simplified significantly (see
Appendix~\ref{appendix:SPproblem} for details) and it turns out
that in order to solve this problem we only need to compute the
order parameter~(\ref{eq:SP1}). The physical meaning of this order
parameter is given by $\Prob^\ell(
S,\hat{S})=\lim_{N\rightarrow\infty}\frac{1}{N}\sum_{i=1}^N\overline{\langle\delta_{S_i^\ell;S}\delta_{\hat{S}_i^\ell;\hat
S}\rangle\vert_{S^I}}$, i.e. the disorder-averaged joint
probability of sites in the two systems.  The single-site
effective measure~(\ref{eq:M}) also benefits from the
simplification; in particular, if we integrate out the continuous
variables in~(\ref{eq:M}) we are led to the expression
%
%
\begin{eqnarray}
M_n[ S^L,\hat{S}^L,\ldots, S^0,\hat{S}^0]&=&\delta_{S^0;S^I_n}\delta_{\hat{S}^0;S^0}\prod_{\ell=0}^{L-1}\Bigg\{\sum_{\{S_j,\hat{S}_j\}}\prod_{j=1}^{k}\left[\Prob^\ell( S_j,\hat{S}_j)\right]\label{eq:effpath}\\
&&\left\langle\times \frac{\rme^{\beta  S^{\ell+1}\alpha(\{S_j\})}}{2\cosh\beta[\alpha(\{S_j\})]}\frac{\rme^{\hat{\beta} \hat{S}^{\ell+1}\alpha(\{\hat{S}_j\})}}{2\cosh\hat{\beta}[\alpha(\{\hat{S}_j\})]}\right\rangle_{\alpha}\Bigg\}.\nonumber
\end{eqnarray}
%
%
Using equation~(\ref{eq:effpath}) the macroscopic
observables~(\ref{def:observ}) can be easily computed from the
joint probability distribution~(\ref{eq:SP1}), resulting in the
set of equations
%
    %
\begin{eqnarray}
&&m(\ell+1)=\sum_{\{S_j\}}\prod_{j=1}^{k}\left[\frac{1}{2}\{1+ S_jm(\ell)\}\right]\left\langle\tanh[\beta \alpha(S_1,\ldots,S_k)]\right\rangle_{\alpha}\label{eq:m}\\
&&C(\ell+1)=\sum_{\{S_j,\hat{S}_j\}}\prod_{j=1}^{k}\left[\frac{1}{4}\{1+ S_j m(\ell)+\hat{S}_j \hat m(\ell)+ S_j\hat  S_j C(\ell)\}\right]\label{eq:Cl}\\
&&~~~~~~~~~~~~~\times\left\langle\tanh[\beta \alpha(S_1,\ldots,S_k)]\tanh[\hat\beta \alpha(\hat{S}_1,\ldots,\hat{S}_k)]\right\rangle_{\alpha}.\nonumber
\end{eqnarray}
%
%
where the magnetization $\hat{m}(\ell)$ is computed by a similar
equation to~(\ref{eq:m}) but with $\beta\rightarrow\hat{\beta}$.
The initial conditions for the above system of equations are given
by $m(0)=\hat{m}(0)=\frac{1}{\vert S^I\vert}\sum_{S\in S^I}
\Spin,~C(0)=1$.

The connectivity profile considered in our model leads to a simple
mean-field theory, where the macroscopic behaviors of the two
copies of the same system is completely determined by the set of
observables $\{m(\ell),\hat{m}(\ell),C(\ell)\}$; which relate to
the order parameter~(\ref{eq:SP1}) via $\Prob^\ell(
S,\hat{S})=\frac{1}{4}(1+ S m(\ell)+\hat{S}\hat m(\ell)+ S\hat{S}
C(\ell))$, while the single system behavior is described by
$\{m(\ell)\} $. Furthermore, since
$\langle\prod_j\Spin^\ell_{i_j}\rangle\!\rightarrow\!\prod_j\langle\Spin^\ell_{i_j}\rangle$
for finite $j$, when $N\rightarrow\infty$ (this can be shown by the direct computation via (\ref{eq:GF})) the spins on
layer $\ell$ are uncorrelated. The reason for this behavior is
that in our model, the site $(i,\ell)$ is a root of a full $k$-ary
tree growing from the input layer $\ell=0$, which in turn points
to the input set $S^I$. The loops in the circuit are rare, so that
trees can be regarded as random Boolean formulas and when
presented with the input, operate independently of each other. The
output of a typical formula at layer $\ell$ is determined by the
probability $\Prob^\ell(S)$.

The overlap order parameter of equation~(\ref{def:observ})
is related to the normalized Hamming distance $D(\ell)$  between
the states $\vecSpin^\ell$ and $\vecSpinhat^\ell$ via the identity
$D(\ell)=\frac{1}{2}(1-C(\ell))$. This allows one to define the
order parameter
$\Delta(\ell)=\lim_{\beta,\hat{\beta}\rightarrow\infty}\frac{1}{2}(1-C(\ell))$,
used to probe sensitivity of the circuit with respect to its
input, an indication to the complexity of the functions
represented by the given circuit. The Hamming distance $D(\ell)$
is also related to the probability
$\Prob(\Spin_i^\ell\neq\Spinhat_i^\ell)$ and facilitates the
estimate of the error probability $\delta(\ell)$ on the $\ell$-th
layer of a noisy circuit. More specifically, we define this error
probability
$\delta(\ell)=\max_{S^I}\lim_{\hat{\beta}\rightarrow\infty}\frac{1}{2}(1-C(\ell))$,
comparing the maximal error between the noisy and noiseless
version of the same circuit with respect to all possible inputs.
Obviously, in the absence of noise ($\beta\rightarrow\infty$) one trivially obtains $\delta(\ell)=0$ for all $\ell$.

\section{Results\label{section:Results}}
In this section we apply equations (\ref{eq:m},\ref{eq:Cl}) to the formulae constructed from a particular single gate $\alpha$ and for a given distribution over gates $P(\alpha)$.

\subsection{MAJ-$k$ gate\label{subsection:MAJ-k}}

\subsubsection{Critical behavior}
\begin{figure}[t]
\vspace{-15mm}
\setlength{\unitlength}{1.4mm}
\begin{picture}(60,55)
\put(5,0){\epsfysize=45\unitlength\epsfbox{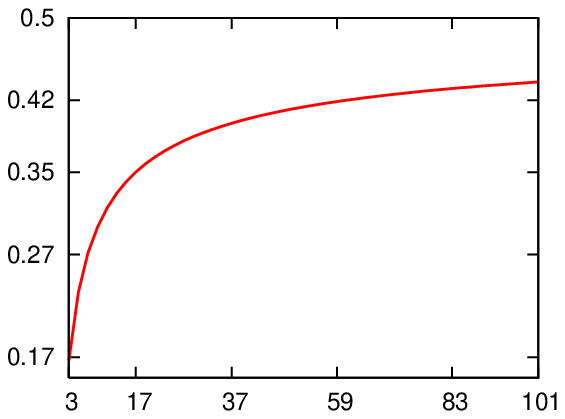}}

\put(5,30){$\epsilon^*$}\put(30,20){$m\!\neq\!0\;(\delta\!<\!^1\!/_2)$}\put(20,35){$m\!=\!0\;(\delta\!=\!^1\!/_2)$}
\put(40,-1){$k$}
\end{picture}
\caption{(Color online) Phase diagram of a circuit based on MAJ-$k$ gates.}
\label{fig:1}
\end{figure}
In this section our choice of $\alpha$ is a universal majority gate with $k$ inputs (MAJ-$k$). The reasons for choosing this gate are twofold. Firstly, it was proved in~\cite{Hajek:MTN},~\cite{Evans:MTNK} that the majority gate is optimal for the noisy computation in formulae. Secondly, formulae constructed from the majority gates can in principle compute any Boolean function~\cite{Savicky}. A convenient representation of the MAJ-$k$ gate is given by the identity $\text{MAJ}(S_1,\ldots,S_k)=\sgn[\sum_{j=1}^k S_j]$, where $k$ is odd. Inserting this identity into the equations (\ref{eq:m}) and  (\ref{eq:Cl}) and computing the spin averages leads to
\begin{eqnarray}
&&m(\ell+1)=(1-2\epsilon)\sum_{n=0}^{k}\binom{k}{n}\left[\frac{1+m(\ell)}{2}\right]^n\left[\frac{1-m(\ell)}{2}\right]^{k-n}\!\! \sgn\left[2n-k\right]\label{eq:m-maj-k}\\
&&C(\ell+1)=(1-2\epsilon)\!\!\sum_{k_1+..+k_4=k}\frac{k!}{k_1!\times..\times k_4!}\;P^{k_1}(-1,-1)\; P^{k_2}(1,-1)\label{eq:C-maj-k}\\
&&~~~~~~~~~~~~~\times P^{k_3}(-1,1)\; P^{k_4}(1,1)\sgn\left[k_1\!-\!k_2\!+\!k_3\!-\!k_4\right]\sgn\left[k_1\!+\!k_2\!-\!k_3\!-\!k_4\right]\nonumber
\end{eqnarray}
where $P(S,\hat S)=\frac{1}{4}(1+S m(\ell)+\hat S\hat m(\ell)+S\hat S C(\ell))$. To obtain equations~(\ref{eq:m-maj-k},\ref{eq:C-maj-k}) we have used the identity $\tanh\beta=1-2\epsilon$ which relates the gate-error $\epsilon$ to the inverse temperature $\beta$. Also, we have taken the limit $\hat\beta\rightarrow\infty$ ($\hat\epsilon=0$) to compare noisy circuit outputs to their noiseless counterparts later on.

For now we concentrate only on  equation (\ref{eq:m-maj-k}) which describes the evolution of magnetization from layer to layer. The point $m(\infty)=0$ is always a stationary solution of this equation, i.e. $m(\ell+1)=m(\ell)=m(\infty)$.  Expanding equation (\ref{eq:m-maj-k}) around this stationary solution gives the critical noise value $\epsilon^*(k)=1/2-2^{k-2}/k\binom{k-1}{(k-1)/2}$, identical to the results of~\cite{Hajek:MTN,Evans:MTNK}; at the critical noise the asymptotic solution $m(\infty)=0$ becomes unstable and two stable solutions $\pm m(\infty)$ (e.g.,for $k=3$ we find $m(\infty)=\pm\sqrt{\frac{1-6\epsilon}{1-2\epsilon}}$~\cite{noisyBoolean}) emerge. In the case when $\epsilon>\epsilon^*(k)$ the magnetization $m(\ell)$ decays to $0$ when $\ell\rightarrow\infty$. For the $\epsilon<\epsilon^*(k)$ we have $\lim_{\ell\rightarrow\infty}m(\ell)=\pm m(\infty)$ where the positive and negative stationary solutions correspond to the positive and negative initial magnetizations $m(0)=\frac{1}{\vert S^I\vert}\sum_{S\in S^I} \Spin$, respectively. Thus the critical noise level  $\epsilon^*(k)$ separates the unordered phase of the system from ordered one (see Figure \ref{fig:1}).

The relation between the new stable solutions and the reliability of the computation follows from the ability to preserve one bit of information presented at the input, by setting $S^I=\{S\}$;  the phase transition observed in equation~(\ref{eq:m-maj-k}) implies that the circuit can preserve one bit of information for arbitrarily many layers only when $\epsilon<\epsilon^*(k)$. The probability of an error $\Prob^\ell(-\Spin)=\frac{1}{2}(1-\Spin m(\ell))$ is a measure of how well this one bit  is preserved after passing through $\ell$  layers. A complicated computational task may require significant number of layers, hence only relatively simple operations can be performed  by the circuit reliably when $\epsilon>\epsilon^*(k)$.

Now we turn to equation~(\ref{eq:C-maj-k}) which describes evolution of the overlap between two systems. The initial conditions are the same for both systems, so we have $m(0)=\hat m(0)$ and $C(0)=1$. The magnetization in the noisy system ($\epsilon<\epsilon^*(k)$) converges to $\pm m(\infty)$ and for the noiseless system we have  $\hat m(\infty)=\pm1$, depending on the sign of $m(0)$. Inserting these stationary points into the equation (\ref{eq:C-maj-k}) results in $C(\infty)=\pm m(\infty)$. The overlap $C(\infty)$ relates to the probability of error $\delta(\infty)=\frac{1}{2}(1-C(\infty))$. Thus the error $\delta(\infty)$ is bounded below $1/2$ only when $\epsilon<\epsilon^*(k)$.

\subsubsection{Boolean functions generated\label{subsection:functions}}
The analysis of equation (\ref{eq:m-maj-k}) can also reveal the type of Boolean functions generated in the layered growth process.  In particular, in the noiseless case ($\epsilon=0$) the stationary solutions of this equation are given by $m(\infty)=1$ and $m(\infty)=-1$ which correspond to the initial conditions  $m(0)>0$ and $m(0)<0$ respectively. For $m(0)=0$ we obtain $m(\infty)=0$. Each site in our model can be associated with an output of the formula that computes some Boolean function. The average formula on layer $\ell$ provides outputs $\Spin$ with probability $\Prob^\ell(\Spin)=\frac{1}{2}(1-\Spin m(\ell))$. This suggests that for the noiseless case $\epsilon=0$ the average formula on layer $\ell$ converges to a random Boolean function
\begin{eqnarray}
F= \left\{
\begin{array}{l l}
  +1 & \quad \mbox{if $m(0)>0$}\\
  \pm1 \quad\mbox{with prob. $^1/_2$}& \quad \mbox{if $m(0)=0$}\label{eq:F}\\
  -1& \quad \mbox{if $m(0)<0$}\\
\end{array} \right.
\end{eqnarray}
where $m(0)=\frac{1}{\vert S^I\vert}\sum_{S\in S^I} \Spin$. This means that depending on the initial conditions the formulae converge to a \emph{single} Boolean function or to the \emph{uniform distribution} over some set of functions.  For example, if we take $S^I=\{-1,-S_1^I,-S_2^I\}$ then  the formulae converge to the NAND function. Taking $S^I=\{-S_1^I,-S_2^I\}$, on the other hand, gives us uniform distribution over the inverse functions $-S_1^I$ and $-S_2^I$ as follows from the majority property of the gate. In general, when $m(0)=0$, is difficult to say if the formulae compute \emph{all} Boolean functions in the set~(\ref{eq:F}) or only the subset of these functions.
However, this result~(\ref{eq:F}) is consistent with the study of Savick\'{y}~\cite{Savicky} where the majority gate forms the basis of the growth process that generates random Boolean formulae. In particular, it has been shown that when  $S^I=\{-1,1,S_1^I,\ldots,S_n^I,-S_1^I,\ldots,-S_n^I\}$ the formulae in the stationary state of the process compute all Boolean functions of $n$ variables. Furthermore, the equation~(\ref{eq:F}) is also consistent with results reported elsewhere~\cite{Brodsky} where the same growth process is considered for various initial conditions $S^I$. In particular, for $S^I=\{-1,1,S_1^I,\ldots,S_n^I\}$ the formulae converge to the MAJ-$n$ function when $n$ is odd and to the uniform distribution over so-called slice functions when $n$ is even \cite{Brodsky}. The same happens when the constants $\{-1,1\}$ are removed from the set $S^I$ \cite{Brodsky}. The result of equation~(\ref{eq:F}) can be written in a more compact form $F=\sgn[\sum_{i=1}^n\theta_i\Spin_i^I+\theta_0]$, where $\theta_i\in\Z$, using the definition: $\sgn[0]=\pm1$ with probability $1/2$. If $\sum_{i=1}^n\theta_i\Spin_i^I+\theta_0\neq0$ for $\forall\;(S_1^I,\ldots,S_n^I)\in\{-1,1\}^n$ then all formulae in the circuit converge to a single \emph{linear threshold function} which can compute any linearly separable Boolean function~\cite{Minsky}.

\subsubsection{Sensitivity of the generated functions\label{subsection:sens}}
\begin{figure}[t]
\vspace{-15mm}
\setlength{\unitlength}{1.4mm}
\begin{picture}(110,55)
\put(-7,0){\epsfysize=45\unitlength\epsfbox{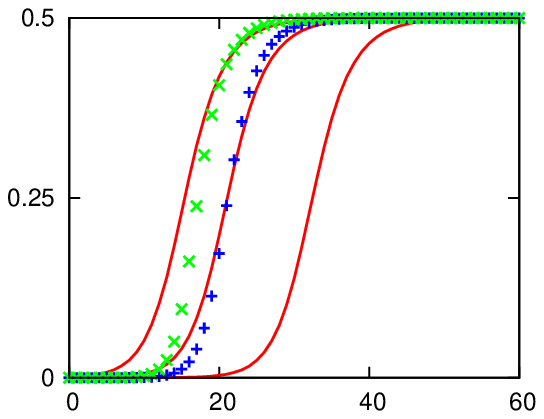}}
\put(30,-1){$\ell$}
\put(-2,25){$\Delta$}\put(10,38){$(a)$}\put(70,38){$(b)$}
\put( 52,  0){\epsfysize=45\unitlength\epsfbox{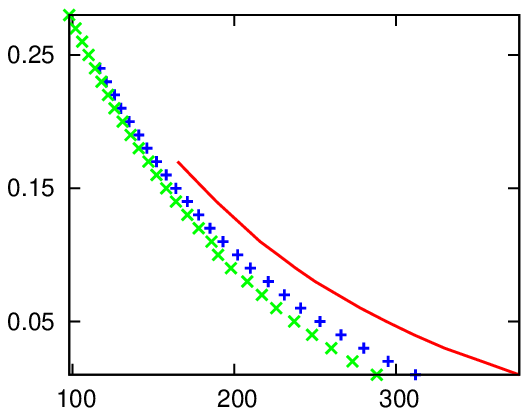}}
\put(57,25){$\epsilon$}\put(87,25){$\delta=1/2$}\put(73,10){$\delta<1/2$}
\put(90,-1){$L$}
\end{picture}
\caption{(Color online) (a) Evolution of the Hamming distance as a function of the layer $\ell$ for $k=3$ (solid line) with input mismatch $\Delta(0)=10^{-3},10^{-4},10^{-6}$ (left to right) and for $k=5$ ($+$) and $k=7$ ($\times$) with $\Delta(0)=10^{-6}$. (b) Phase boundaries for gate-noise $\epsilon$ at layer $L$ when $k=3,5,7$, using the same symbols.}
\label{fig:2}
\end{figure}

We will now turn to equation~(\ref{eq:C-maj-k}) when $m(0)=\hat m(0)=0$ and $C(0)=1$. For the noiseless case $\epsilon=0$ the stationary solution of this equation is given by  $C(\infty)=1$. This solution, however, is unstable and a small perturbation to the initial state $C(0)=1$ leads to the stationary state $C(\infty)=0$, which is stable. This implies that the circuit is very sensitive to its input when $m(0)=0$. In particular, the Hamming distance $\Delta(\ell)=\frac{1}{2}(1-C(\ell))$ increases for small perturbations to $\Delta(0)$ (see Figures \ref{fig:2} (a)), i.e. a small perturbation to the input is amplified by the circuit. This in turn means that when $\epsilon>0$ the circuit also amplifies  \emph{the noise-perturbation} and the error $\delta(\ell)$ is growing. The error, however, can be kept under control for many layers by making $\epsilon$ sufficiently small (see Figure~\ref{fig:2} (b)).

\subsubsection{Dynamics\label{subsection:dynamics}} %
We will now examine how the computation in the circuit proceeds from layer to layer. As an example we take $S^I=\{S_1,\ldots,S_{11}\}$, i.e. for $\epsilon=0$ the circuit computes the MAJ-$11$ function, and study the evolution of magnetization $m(\ell)$ and error $\delta(\ell)$. The initial magnetization $\vert m(0)\vert$ takes its values from the set $\{1,9/11,7/11,5/11,3/11,1/11\}$ for this choice of $S^I$. The input with the smallest magnetization in this set is very important. On the one hand, when $\epsilon=0$ the initial state with the smallest $\vert m(0)\vert$ is also the furthest from the  stationary state $\vert m(\infty)\vert=1$. So it will take for the magnetization $m(\ell)$ the largest number of layers to converge for this input. On the other hand, when $\epsilon>0$ the input with the smallest $\vert m(0)\vert$ is more likely to be destroyed by the noise. For these reasons, in what follows we study the evolution of the magnetization and errors only for $m(0)=1/11$.

\begin{figure}[t]
\vspace{-0mm}
\setlength{\unitlength}{1.4mm}
\begin{picture}(100,120)
\put( 53,  80){\epsfysize=40\unitlength\epsfbox{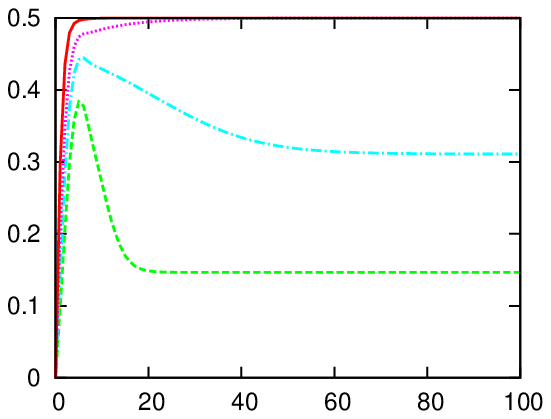}}
\put(0,80){\epsfysize=40\unitlength\epsfbox{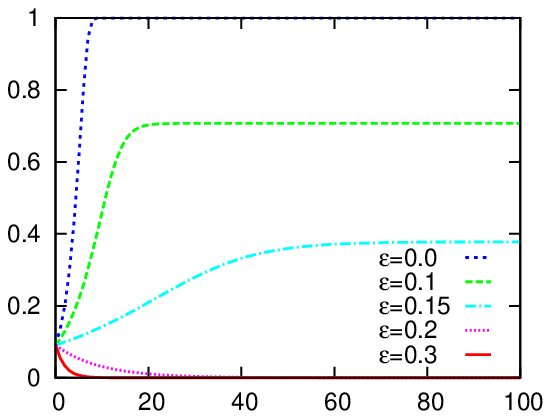}}
\put( 53,  40){\epsfysize=40\unitlength\epsfbox{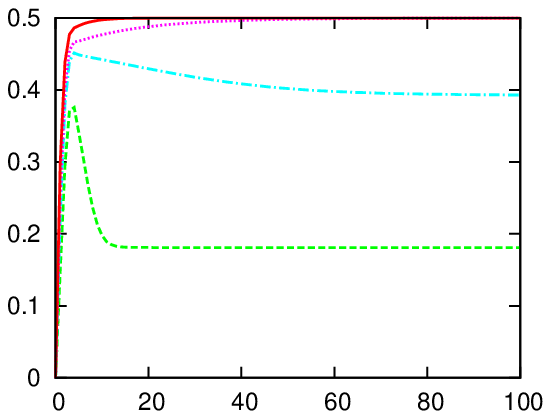}}
\put(0,40){\epsfysize=40\unitlength\epsfbox{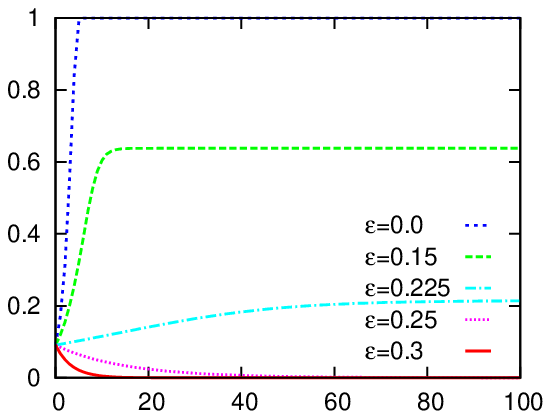}}
\put( 53,  0){\epsfysize=40\unitlength\epsfbox{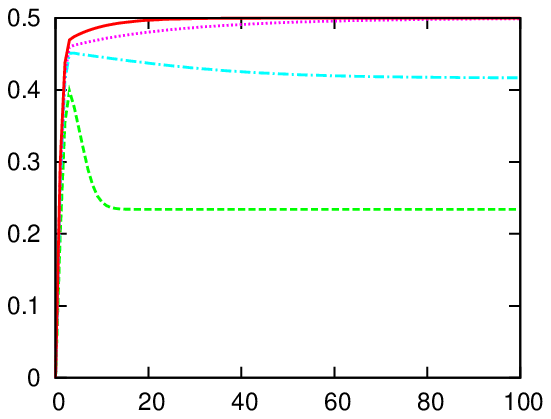}}
\put(0,0){\epsfysize=40\unitlength\epsfbox{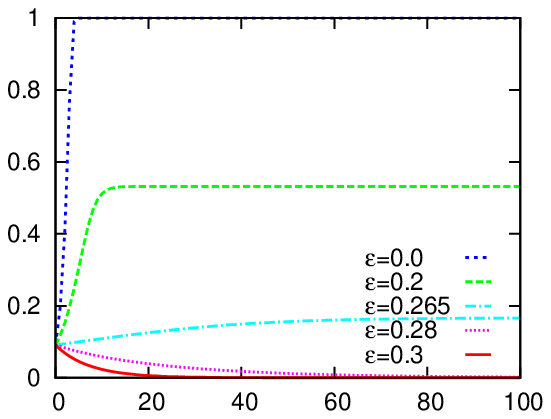}}
\put(57,100){$\delta$}
\put(3,100){$m$}
\put(32,-1){$\ell$}\put(85,-1){$\ell$}
\put(32,120){$(a)$}\put(85,120){$(b)$}
\put(25,110){$k=3$}
\put(25,70){$k=5$}
\put(25,30){$k=7$}
\end{picture}
\caption{(Color online) 
Evolution of magnetization (a) and error (b) in MAJ-$k$ formulae for  $k=3,5,7$ ($\epsilon^*(3)\approx0.167,\epsilon^*(5)\approx0.233,\epsilon^*(7)\approx0.271$) and different noise ($\epsilon$) values.}
\label{fig:4}
\end{figure}

In Figure \ref{fig:4} (a) we examine how the magnetization $m(\ell)$ evolves from layer to layer in circuits with MAJ-$3$, MAJ-$5$ and MAJ-$7$ gates. We observe that when $\epsilon=0$ the magnetization converges to its stationary value $m(\infty)=1$ relatively quickly. Since we use $m(0)=1/11$ the convergence to $m(\ell)=1$ indicates that all formulae in the circuit compute the MAJ-$11$ function. For noise values $\epsilon>0$ the speed of convergence is decreasing as $\epsilon$ increases  and becomes very slow as we approach $\epsilon^*(k)$. In general, increasing $k$ ($0\leq\epsilon<\epsilon^*(k)$)  leads to a reduction in relaxation times because of the inequality $F^{k+2}_\epsilon[m]\geq F^{k}_\epsilon[m]\geq m$, where $F^{k}_\epsilon[m]$ is the right hand side of equation (\ref{eq:m-maj-k}). Finally, when we increase the noise level to $\epsilon\gg\epsilon^*(k)$ the magnetization relaxes to its stationary $0$ value exponentially fast.

Figure~\ref{fig:4}(b) shows the evolution of the error $\delta(\ell)$. In the region $0<\epsilon<\epsilon^*(k)$ we observe two distinct stages in the dynamics. Initially, the error is increasing until it reaches its maximum value. Note that this happens before the MAJ-$11$ function is computed exactly when $\epsilon=0$ (see Figure \ref{fig:4} (a)). Also, the location of this maximum is only weakly affected by noise. These two observations suggest that initially the inputs to the gates are very inhomogeneous which leads to the amplification of noise. After the error reaches its maximum value the inputs become more and more homogeneous leading to the suppression of noise and as a result the error decreases until it eventually becomes stationary. As we approach the critical boundary $\epsilon^*(k)$ the number of layers needed for the error to equilibrate  increases. The dynamic behavior of the error changes from the non-monotonically increasing to the monotonically increasing when we approach the critical boundary $\epsilon^*(k)$ from below. The evolution of error becomes strictly monotonic when $\epsilon\gg\epsilon^*(k)$ and in this region the error relaxes to its stationary value $1/2$ exponentially fast.

This example is highly representative of the situation when all formulae in the circuit converge to a single Boolean function. Here we can tell exactly how many layers it takes for the circuit to compute this function when $\epsilon=0$. This number is given by $L$ such that $m(L)\approx m(\infty)$ starting with the smallest initial magnetization $m(0)$ induced by the inputs $S^I$. Obviously, adding more layers to the circuit with $L$ layers will not affect the computation when $\epsilon=0$. However, in the region $0<\epsilon<\epsilon^*(k)$ we can reduce the error $\delta(\ell)$ by adding more layers. This is not true for all $\epsilon$ and $m(0)$. The error can be reduced only when $\epsilon<\epsilon^0(k)$, where $\epsilon^0(k)$ is the solution of $m(0)=F_{\epsilon^0(k)}^k[m(0)]$ for a given initial magnetization $m(0)$. At $\epsilon=\epsilon^0(k)$ the dynamic behavior of $m(\ell)$ changes from the monotonically increasing (when $\epsilon<\epsilon^0(k)$) to monotonically decreasing (when $\epsilon>\epsilon^0(k)$). Respectively, the dynamic behavior of the error $\delta(\ell)$ changes from the non-monotonically increasing to the monotonically increasing. Only in the former regime one can reduce the error $\delta(\ell)$ by adding more layers to the circuit. However, this strategy fails for inputs with $m(0)=0$ and the circuit computes more than one Boolean function. For $m(0)=0$ the probability of error $\delta(\ell)$  increases towards its stationary value $\delta(\infty)=1/2$ ($m(\ell)=0$) even when $\epsilon<\epsilon^*(k)$. The error $\delta(\ell)$ can be bounded away from $1/2$ only by reducing the gate-error $\epsilon$, which depends on the formula depth $L$ (Figure~\ref{fig:2} (b)).

\subsubsection{Convergence rates\label{subsection:rates}}
In this section we study convergence rates at $\epsilon=0$ and $\epsilon=\epsilon^*(k)\pm\Delta\epsilon$ regions of the phase diagram plotted in Figure \ref{fig:1}. The former allows us to estimate the number of layers in a formula, which is directly related to its \emph{size}, when all formulae in the circuit converge to a single Boolean function. The latter probes the regime where the computation is expected to be very slow, but the error can be still reduced by adding more layers to the circuit.

\begin{figure}[t]
\vspace{-15mm}
\setlength{\unitlength}{1.4mm}
\begin{picture}(60,55)
\put( 5,  0){\epsfysize=45\unitlength\epsfbox{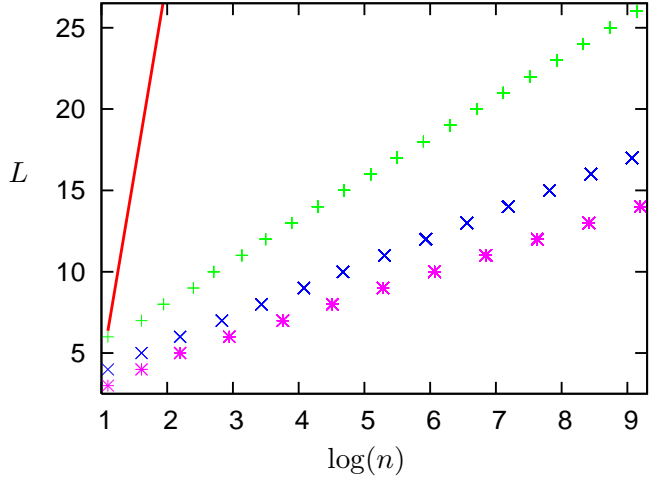}}
\put(5,25){$L$}
\put(35,-2){$\log(n)$}
\end{picture}
\caption{(Color online) Number of layers in the noiseless MAJ-$k$ based circuit computing MAJ-$n$ function. Theoretically obtained data-points are represented by the symbols $+$ ($k\!=\!3$) , $\times$ ($k\!=\!5$)  and $*$ ($k\!=\!7$). Slopes of the respective straight lines (not shown) fitted to the data  are $2.473$, $1.595$ and $1.282$, respectively. The straight line corresponds to the bound $k2^k\log(n)$, derived in~\cite{Brodsky}, plotted here for $k\!=\!3$ only.}
\label{fig:10}
\end{figure}

Firstly we study the rate of convergence when $\epsilon=0$ and $m(0)=1/n$, where $n\in\mathbb{N}$ is odd, and the MAJ-$k$ based circuit computes MAJ-$n$ function. In general, we find that the number of layers needed for the magnetization to converge scales as $O(f(k)\log (n))$. This rate of convergence is consistent with  rigorous results~\cite{Brodsky} for the growth process defined by Savick\'{y}~\cite{Savicky}. However, the \emph{worst case bound} $f(k)$ derived in~\cite{Brodsky} grows as $k2^k$ with the gate in-degree $k$, while in our study we find that $f(k)$ is \emph{decreasing} with increasing $k$ (see Figure~\ref{fig:10} ). Furthermore, this result holds not only for MAJ-$n$, but for \emph{any} linear threshold function (with integer weights) computed by the MAJ-$k$ circuit. It is natural to expect that when $k\rightarrow\infty$ the function $f(k)$ is vanishing and the numbers of layers $L$ in the circuit approaches $1$. The discrepancy in the asymptotic behavior of the worst case bounds~\cite{Brodsky} and the \emph{typical} asymptotic behavior observed in our work is due to the \emph{average topology } considered here, which turns out to be more realistic.
	
Secondly, we study convergence rate for $\epsilon=\epsilon^*(k)\pm\Delta\epsilon$, where $0<\Delta\epsilon\ll1$.  Very close to the phase boundary $\epsilon^*(k)$ the differences $m(\ell+1)-m(\ell)$ are very small and the difference equation (\ref{eq:m-maj-k}) can be well approximated by a differential equation. For $k=3$ the differential equation reduces to
$\frac{\mathrm{d}}{\mathrm{d}\ell}m(\ell)\!=\!-m(\ell)\!+\!\frac{1}{2}(1\!-\!2\epsilon)[3m(\ell)\! -\! m^3(\ell)]$, where $\ell$ is continuous, which can be solved exactly. The solution is given by $m^2(\ell)=\left\{  \left[ \frac{1}{m^2(0)}\!-\!\frac {1\!-\!2\epsilon}{1\!-\!6\epsilon}
 \right] \rme^{- ( 1\!-\!6\epsilon)\, \ell}\!+\!\frac {1\!-\!2\epsilon}
{1\!-\!6\epsilon} \right\}^{-1}$. This approach is only accurate in the region $\epsilon=1/6\pm\Delta\epsilon$, where it gives us the asymptotic form $\vert m(\ell)\!-\!m(\infty)\vert\approx\rme^{-\gamma(3)\Delta\epsilon\ell}$. The $\gamma(3)$ coefficient equals $3$ in the paramagnetic region and $6$ in the ferromagnetic region. Thus the convergence to the asymptotic solution $m(\infty)=0$ is slower than to the stationary solutions $m(\infty)=\pm\sqrt{\frac{1-6\epsilon}{1-2\epsilon}}$ for $k=3$. The differential version of the difference equation (\ref{eq:m-maj-k}) is difficult to solve analytically when $k>3$ and for these values of $k$ we will use a different method to estimate the convergence rate. This method relies on the fact that
\begin{eqnarray}
\textrm{const}\times\left\vert \left\{\frac{\rmd}{\rmd m}F_{\epsilon}^{k}[m]\right\}_{m=m(\infty)}\right\vert^{\ell}\leq\vert m(\ell)-m(\infty)\vert\label{eq:ineq},
\end{eqnarray}
where $F_{\epsilon}^k[m]$ is the right hand side of equation (\ref{eq:m-maj-k}), i.e. the distance $\vert m(\ell)-m(\infty)\vert$ for an arbitrary point $m(\ell)$ is always greater than the distance for $m(\ell)=m(\infty)+\Delta m(\ell)$. The lower bound (\ref{eq:ineq}) can be made into an upper bound by choosing an appropriate constant~\cite{Ortega}. Computing the left hand side in (\ref{eq:ineq}) for the MAJ-$k$ circuit with $\epsilon=\epsilon^*(k)\pm\Delta\epsilon$ leads to the result
\begin{eqnarray}
\left\{\left[1-\frac{2\Delta\epsilon}{1-2\epsilon^*(k)}\right](1-m^2(\infty))^{(k-1)/2}\right\}^{\ell}\leq\vert m(\ell)-m(\infty)\vert.\label{eq:ineqMAJk}
\end{eqnarray}
Since $\left[1-\frac{2\Delta\epsilon}{1-2\epsilon^*(k)}\right](1-m^2(\infty))^{(k-1)/2}\leq\left[1-\frac{2\Delta\epsilon}{1-2\epsilon^*(k)}\right]$, the convergence rate in the paramagnetic region ($\epsilon=\epsilon^*(k)+\Delta\epsilon$) is slower than in the ferromagnetic one ($\epsilon=\epsilon^*(k)-\Delta\epsilon$). The latter is due to the amplification of thermal fluctuations which are only suppressed for $\ell\rightarrow\infty$. For large $k$ the critical noise $\epsilon^*(k)$ can be approximated by $\epsilon^*(k)\approx\frac{1}{2}(1-\frac{\sqrt{\pi}}{\sqrt{2k}})$. Inserting this into the equation~(\ref{eq:ineqMAJk}) for $m(\infty)=0$ gives the asymptotic form
 \begin{eqnarray}
\vert m(\ell)-m(\infty)\vert\approx\rme^{-O(k^0)\sqrt{k}\Delta\epsilon\ell},\label{eq:asymptotic}
\end{eqnarray}
from which is clear that increasing $k$ speeds up the convergence in both paramagnetic and ferromagnetic regimes.
\subsubsection{Hard noise}
The model studied so far can be regarded as a model of computation where errors result from single-event upsets (soft noise). In real integrated circuits~\cite{Borkar}, the imperfections introduced into the circuit during the production process are an additional source of permanent errors (hard noise).

A natural way to introduce hard noise into our model is to define \emph{quenched} random variables $\{\xi_i^\ell\}$, where  $\Prob(\xi_i^\ell)=p\delta_{\xi_i^\ell;\minus1}+(1-p)\delta_{\xi_i^\ell;1}$, which can invert the gate output $\alpha_i^\ell$ permanently. Using transformation $\alpha_i^\ell\rightarrow\xi_i^\ell\alpha_i^\ell$ in  equation~(\ref{eq:micro}) and following the steps of calculation in section \ref{section:Method}, we find that the inclusion of hard noise in our model leads to $(1-2\epsilon)\rightarrow(1-2p)(1-2\epsilon)$ in equations~(\ref{eq:m-maj-k}) and (\ref{eq:C-maj-k}). As a result, the effect of quenched noise is to reduce the critical noise $\epsilon^*(k)$. In particular, the new critical noise value is given by $\epsilon^*(k,p)=\frac{1}{2}-\frac{2^{k-2}}{(1-2p)k\binom{k-1}{(k-1)/2}}$ when $0\leq p<\epsilon^*(k,0)$ and $\epsilon^*(k,p)=0$ when $p\geq\epsilon^*(k,0)$.

The hard noise can also be introduced by making a fraction of gates insensitive to the inputs, i.e. gates that produce constants. In particular, by taking $P(\alpha)=p_0\delta_{\alpha;\textrm{MAJ-k}}+p_-\delta_{\alpha;\minus1}+p_+\delta_{\alpha;1}$, where $p_{\pm}$ are the probabilities of constant $\pm1$ outputs and $p_0,p_-,p_+\geq0$ with $p_0+p_-+p_+=1$, equation~(\ref{eq:m})  for $p_+-p_-=0$ results in  $\epsilon^*(k,p_0)=\frac{1}{2}-\frac{2^{k-2}}{p_0k\binom{k-1}{(k-1)/2}}$ when $0\leq (1-p_0)/2<\epsilon^*(k,p_0=0)$ and $\epsilon^*(k,p_0)=0$ when $(1-p_0)/2\geq\epsilon^*(k,p_0=0)$. So the introduction of constant gates reduces the critical noise value $\epsilon^*(k)$ when $p_+=p_-$ by effectively reducing the number of active gates. For $p_+-p_-\neq0$ the effect of hard noise is more drastic. For $\epsilon=0$ the circuit, irrespective of its input, is biased towards one of its outputs $\pm1$, depending on the value of $p_+-p_-$.

\subsubsection{Threshold noise} 
\begin{figure}[t]
\vspace{-15mm}
\setlength{\unitlength}{1.4mm}
\begin{picture}(60,55)
\put(5,0){\epsfysize=45\unitlength\epsfbox{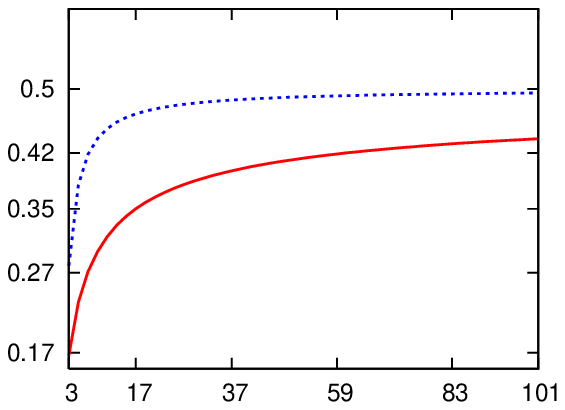}}
\put(5,30){$\epsilon^*$}\put(30,20){$m\!\neq\!0\;(\delta\!<\!^1\!/_2)$}\put(20,35){$m\!=\!0\;(\delta\!=\!^1\!/_2)$}
\put(40,-1){$k$}
\end{picture}
\caption{(Color online) The critical noise level $\epsilon^*$ as a function of $k$ for the perceptron (dotted line) and MAJ-$k$ (solid line) circuits.}
\label{fig:11}
\end{figure}
The MAJ-$k$ function can be seen as a special case of a linear threshold function. The linear threshold functions (or binary perceptrons) are widely used in the modeling of the neuronal activities of the brain such as memory and learning~\cite{TheorOfIP}. The noise in these models is usually introduced via random thresholds. The stochastic process for the simplest model of this class is governed by the algorithm~\cite{TheorOfIP}
\begin{eqnarray}
S_{i}^\ell=\sgn\left[\sum_{j=1}^k S_{i_j}^{\ell-1}+\beta^{-1}\eta_{i}^{\ell-1}\right]\label{def:PercAlg},
\end{eqnarray}
where $\eta_{i}^\ell\in\mathbb{R}$ are independent random variables drawn from the distribution $\Prob(\eta)=\frac{1}{2}[1-\tanh^2(\eta)]$, which generates the state update of individual neurons (on site $i$) at step $\ell$. The stochastic algorithm (\ref{def:PercAlg}) gives rise to the microscopic law
	\begin{eqnarray}
\Prob(S_{i}^\ell\vert S_{i_1}^{\ell-1},\ldots,S_{i_k}^{\ell-1})&=&\frac{\rme^{\beta S_{i}^\ell\sum_{j=1}^k S_{i_j}^{\ell-1}}}{2\cosh[\beta\sum_{j=1}^k S_{i_j}^{\ell-1}]}~,\label{eq:PercMicro}
\end{eqnarray}
which in the limit of $\beta\rightarrow\infty$ corresponds to the noiseless MAJ-$k$ gate. Using the microscopic law (\ref{eq:PercMicro}) in the model of noisy computation (\ref{eq:PathProb}) requires only minor change to the calculations in section~\ref{section:Method} and leads straightforwardly to the corresponding equations for magnetization and overlap
\begin{eqnarray}
&&m(\ell+1)=\sum_{n=0}^{k}\binom{k}{n}\left[\frac{1+m(\ell)}{2}\right]^n\left[\frac{1-m(\ell)}{2}\right]^{k-n}\!\! \tanh[\beta(2n-k)]\label{eq:m-perc-k}\\
&&C(\ell+1)=\sum_{k_1+..+k_4=k}\frac{k!}{k_1!\times..\times k_4!}\;P^{k_1}(-1,-1)\; P^{k_2}(1,-1)\label{eq:C-perc-k}\\
&&~~~~~~~~~~~~~\times P^{k_3}(-1,1)\; P^{k_4}(1,1)\tanh[\beta(k_1\!-\!k_2\!+\!k_3\!-\!k_4)]\sgn\left[k_1\!+\!k_2\!-\!k_3\!-\!k_4\right].\nonumber
\end{eqnarray}
In the limit $\beta\rightarrow\infty$($\epsilon=0$) equations~(\ref{eq:m-perc-k}-\ref{eq:C-perc-k}) and (\ref{eq:m-maj-k}-\ref{eq:C-maj-k}), with $\epsilon=0$, are identical. So that all the results derived for the noiseless MAJ-$k$ circuit are also valid here.

The macroscopic behavior of the model with noisy thresholds, however, is different from the noisy MAJ-$k$ model for any $\epsilon>0$. Analysis of equation~(\ref{eq:m-perc-k}) reveals that the point $m(\infty)=0$ is always a solution of this equation. Expanding equation~(\ref{eq:m-perc-k}) around this point leads to the condition
\begin{eqnarray}
1&=&2^{-k}\sum_{n=0}^{k}\binom{k}{n}\vert2n-k\vert\tanh\beta\vert2n-k\vert\label{eq:PhaseBoundary}
\end{eqnarray}
where $m(\infty)=0$ becomes unstable and two stable $\pm m(\infty)$ solutions emerge. In Figure~\ref{fig:11}, we compare the resulting phase boundary with that of the MAJ-$k$ based circuits. The MAJ-$k$ gate is more resilient to the threshold noise (\ref{def:PercAlg}) than to the flip noise (\ref{def:algorithm}). This is not surprising since the effect of flip noise on the MAJ-$k$ gate  (inverting the gate-output regardless of the input) is more drastic than the effect of threshold noise where gate-outputs $S_i^{\ell}$ corresponding to inputs with high input-magnetization $\vert\sum_{j=1}^k S_{i_j}^{\ell-1}\vert$ are less likely to be flipped. Furthermore, as $k\rightarrow\infty$ the critical noise level $\epsilon^*(k)$ in both models approaches $1/2$ as $1/2-\epsilon^*(k)=O(k^{-\gamma})$, but for the threshold noise model it can be shown that $\gamma=1$ while for the model with flip noise $\gamma=1/2$.

For the noisy threshold model considered here the evolution of magnetization and $\delta$-error is qualitatively similar to the evolution of these order parameters in the flip-noise model. However, we find that the convergence to the stationary state is much quicker in the noisy threshold model. For large $k$ it can be shown that the convergence to all stationary states is dominated by
 \begin{eqnarray}
\vert m(\ell)-m(\infty)\vert\approx\rme^{-O(k^0)k\Delta\epsilon\ell},\label{eq:PercepAsymptotics}
\end{eqnarray}
which is significantly quicker than (\ref{eq:asymptotic}).

Finally, we note that the magnetization equation (\ref{eq:m-perc-k}) is exactly equal to the magnetization equation derived for the parallel dynamics of an Ising ferromagnet on fully asymmetric Bethe lattice~\cite{MimuraAndCoolen,NeriAndBolle} if the layers in the circuit-model are regarded as the time-steps of parallel dynamics. This suggests that the site-time topology (recurrent network) generated by the parallel dynamics is similar to the topology of layered networks considered here when $N\rightarrow\infty$. Based on this observation, which was exploited for instance in~\cite{Derrida}, we expect that the computation performed by the recurrent and layered networks to be the same at least for $m(0)\neq0$. The only differences (if any) can arise from the non-vanishing connected correlations between the different times (layers) in the recurrent network.

\subsection{NAND gate\label{subsection:NAND}}

Here we apply the theory to the formulae constructed by $2$-input universal NAND gates which, according to~\cite{Evans:MTN,Unger}, are optimal for the noisy computation by a $2$-ary Boolean formulae.
The NAND gate can be represented as the linear threshold function $\sgn[-S_1-S_2-1]$.
Using this representation in the magnetization (\ref{eq:m}) and  overlap (\ref{eq:Cl}) equations, with $\hat\beta\rightarrow\infty$ and $\tanh\beta=1\!-\!2\epsilon$, one obtains
\begin{eqnarray}
&&m(\ell+1)=\frac{1}{2}(1-2\epsilon)\left[ (1-m(\ell))^2-2\right]\label{eq:m-NAND}\\
&&C(\ell+1)=\frac{1}{4}(1\!-\!2\epsilon)\big[1\!+\!2m(\ell)\!+\!2\hat m(\ell)\!+\!2C(\ell)\!-\!m^2(\ell)\!+\!2m(\ell)\hat m(\ell)\nonumber\\
&&\!-\!2m(\ell)C(\ell)\!-\!\hat m^2(\ell)\!-\!2\hat m(\ell)C(\ell)\!+\!C^2(\ell)\big]\label{eq:C-NAND}
\end{eqnarray}
where $m(0)=\hat m(0)=\frac{1}{\vert S^I\vert}\sum_{S\in S^I} \Spin$, $C(0)=1$. Here the equation for $\hat m(\ell)$ is identical to equation~(\ref{eq:m-NAND}), but with $\epsilon=0$.
The magnetization equation~(\ref{eq:m-NAND}) admits only one steady state ($m(\infty)=m(\ell+1)=m(\ell)$) solution in the region $m(\infty)\in[-1,1]$. However,
the solution $m(\infty)=1-\frac{1-\sqrt{8\epsilon^2-12\epsilon+5}}{2\epsilon-1}$ to  equation~(\ref{eq:m-NAND}) becomes unstable for the values of noise $\epsilon\leq\epsilon^*=(3-\sqrt{7})/4$ which identifies $\epsilon^*$ as the critical threshold.
Above this threshold the (output) magnetization on layer $\ell\rightarrow\infty$ converges to the value $m(\infty)$, which is independent of the initial (input) magnetization $m(0)$.
Below the threshold $\epsilon^*$ the magnetization $m(\ell)$ oscillates from layer to layer and the properties of this oscillation depends on $m(0)$.
The latter suggests that the circuit performs computation only in the region $\epsilon<\epsilon^*$, in agreement with~\cite{Evans:MTN}.
\begin{figure}[t]
\vspace{-15mm}
\setlength{\unitlength}{1.4mm}
\begin{picture}(60,55)
\put(5,0){\epsfysize=45\unitlength\epsfbox{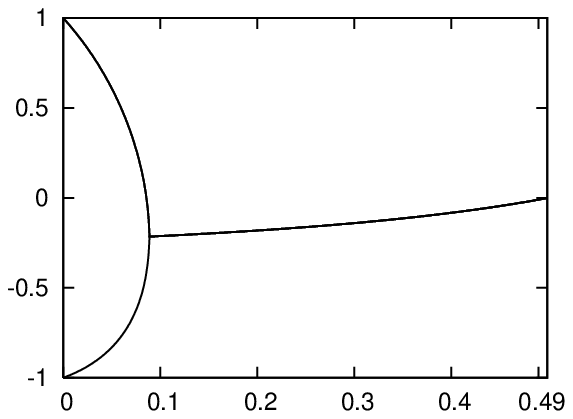}}
\put(20,20.5){\epsfysize=22\unitlength\epsfbox{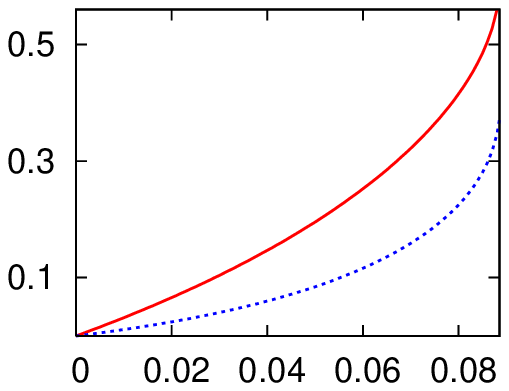}}
\put(40,33){\small{$\delta_+$}}\put(43,30){\small{$\delta_-$}}
\put(5,30){$m$}
\put(40,-1){$\epsilon$}
\end{picture}
\caption{(Color online) Magnetization $m$ and error $\delta$ as a function of gate-noise $\epsilon$ in NAND-gate formulae. The dependence of $\delta_{\pm}$ on the noise level in the range $0\!\leq\!\epsilon\!\leq\!(3\!-\!\sqrt{7})/4$ is shown in the inset. }
\label{fig:12}
\end{figure}

To explore the properties of this computation we follow the evolution of the probability $P_{-}(\ell)=(1\!-\!m(\ell))/2$ . Using this definition in the equation~(\ref{eq:m-NAND}) we obtain the difference equations for $P_{-}(\ell)$ on even layers
\begin{eqnarray}
P_{-}(\ell+2)=1\!-\!\epsilon\!-\!(1\!-\!2\epsilon)(1\!-\!\epsilon)^2+
2(1\!-\!\epsilon)(1\!-\!2\epsilon)^2P^2_{-}(\ell)-(1\!-\!2\epsilon)^3P^4_{-}(\ell)
\label{eq:Prob-NAND}
\end{eqnarray}
where $P_{-}(0)=(1\!-\!m(0))/2$. Equation~(\ref{eq:Prob-NAND}) also describes the evolution of $P_{-}(\ell)$ on odd layers with the initial condition being $P_{-}(1)=\frac{1}{2}(1\!-\!2\epsilon)\left[P^2_{-}(0)\!-\!\frac{1}{2}\right]$.
In the region $P_{-}\in[0,1]$  equation~(\ref{eq:Prob-NAND}) admits three steady state solutions $P_{-}(\infty)=\left\{\frac{1-\sqrt{8\epsilon^2-12\epsilon+5}}{4\epsilon-2}, \frac{-1\pm\sqrt{8\epsilon^2-12\epsilon+1}}{4\epsilon-2}\right\}$.
The first solution becomes unstable at the noise threshold $\epsilon^*=(3-\sqrt{7})/4$ and the other two solutions are stable for the noise values $\epsilon<\epsilon^*$.
Plotting these solutions with respect to the noise $\epsilon$ in terms of the corresponding magnetization variables $m(\infty)$ gives the phase diagram depicted in Figure~\ref{fig:12}.
The stationary solutions of equation~(\ref{eq:Prob-NAND}) also allow one to compute the $\delta$-error which, due to oscillatory behavior of the magnetization $m(\ell)$ in (\ref{eq:m-NAND}), depends on the sign of the output.
In particular the error $\delta(\infty)$ takes its values from the set $\{\delta_-=\frac{-1+\sqrt{8\epsilon^2-12\epsilon+1}}{4\epsilon-2},\delta_+=1-\frac{-1-\sqrt{8\epsilon^2-12\epsilon+1}}{4\epsilon-2}\}$ when $m(0)\notin(2-\sqrt{5},1-\frac{1-\sqrt{8\epsilon^2-12\epsilon+5}}{2\epsilon-1})$ and $\delta(\infty)\in\{1-\delta_+,1-\delta_-\}$ when $m(0)\in(2-\sqrt{5},1-\frac{1-\sqrt{8\epsilon^2-12\epsilon+5}}{2\epsilon-1})$.
The dependence of the error-functions $\delta_{\pm}$ on the gate-noise $\epsilon$ is shown in the inset of Figure~\ref{fig:12}.

For $\epsilon=0$ the basins of attraction of the fixed points $P_-(\infty)\in\{1,0\}$ are given by $P_-(0)\in[1,\frac{\sqrt{5}-1}{2})$ and $P_-(0)\in(\frac{\sqrt{5}-1}{2},0]$ respectively.
Thus for $\epsilon=0$ and $\ell\rightarrow\infty$ the NAND formulae compute the Boolean function
\begin{eqnarray}
F= \left\{
\begin{array}{ll}
  -1 & \quad \mbox{if $m(0)<2-\sqrt{5}$}\label{eq:Fnand}\\
  +1& \quad \mbox{if $m(0)>2-\sqrt{5}$}\\
\end{array} \right. ,
\end{eqnarray}
where $m(0)=\frac{1}{\vert S^I\vert}\sum_{S\in S^I} \Spin$, when $\ell$ is even and its inverse $-F$ when $\ell$ is odd.
In contrast to the result for MAJ-$k$ circuit~(\ref{eq:F}), the variety of functions generated by the NAND formulae is rather limited.
All formulae in the NAND circuit converge to the linear threshold function~(\ref{eq:Fnand})  which, due to the threshold value being equal to $2-\sqrt{5}$, cannot compute all linearly separable Boolean functions.

\begin{figure}[t]
\vspace{-15mm}
\setlength{\unitlength}{1.4mm}
\begin{picture}(110,55)
\put(-10,0){\epsfysize=45\unitlength\epsfbox{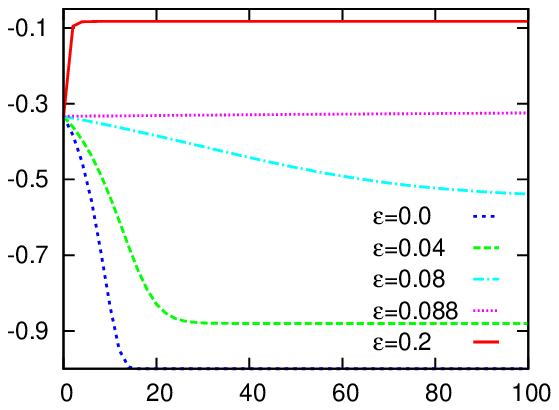}}
\put(25,-1){$\ell$}
\put(-10,25){$m$}
\put( 50,  0){\epsfysize=45\unitlength\epsfbox{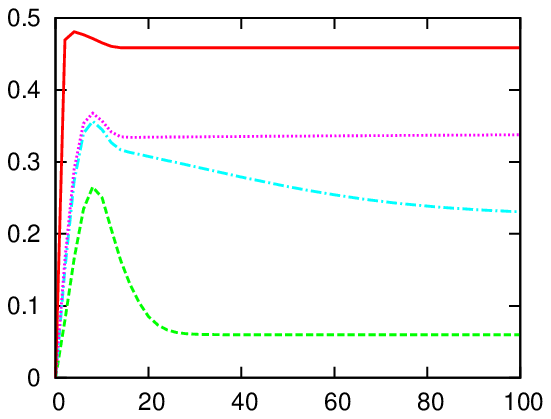}}
\put(53,25){$\delta$}
\put(85,-1){$\ell$}
\end{picture}
\caption{(Color online) Evolution of magnetization and $\delta$-error on even layers in NAND formulae computing MAJ-$3$ function.}
\label{fig:14}
\end{figure}

As an example of a noisy computation by the NAND circuit we consider the input set $S^I=\{S_1,S_2,S_3\}$; the initial magnetization is given by $m(0)=(S_1+S_2+S_3)/3\in\{-1,-1/3,1/3,1\}$ and according to (\ref{eq:Fnand}) the noiseless circuit converges to the MAJ-$3$ Boolean function on even layers and to its dual on odd layers.
In Figure \ref{fig:14} we plot the evolution of magnetization $m$ and $\delta$-error only for $m(0)=-1/3$ where the number of layers $\ell$ needed for the magnetization $m(\ell)$ to converge to its stationary value $-1$ (for $\ell$ even) is maximal.

We observe that for $\epsilon=0$ the magnetization approaches its stationary value $-1$ in approximately $16$ layers, so all the formulae compute MAJ-$3$ function after $16$ layers.
For gate-noise values  $\epsilon>0$, the number of layers $L$ needed for the magnetization to became stationary increases as we increase $\epsilon$ towards its critical value $\epsilon^*$, while the stationary magnetization value $m(\infty)$ decreases ($\vert m(\infty)\vert<1$).
As a result, the stationary values of the $\delta$-error, which is directly related to $m(\infty)$, grow monotonically towards $\delta(\infty)$.
The error grows first then reduces with the addition of layers (see Figure~\ref{fig:14}); this reduction of error is only possible when $\epsilon<\epsilon(0)=\frac{m^2(0)-m(0)-3+\sqrt{-3m^2(0)+4m(0)+8}}{2(m^2(0)-2m(0)-1)}$.
Above the critical noise level $\epsilon^*$ the stationary magnetization $m(\infty)$ is independent of $m(0)$ and the computation becomes unreliable.

Following the method outlined in section~\ref{subsection:MAJ-k}, we obtain the rate of convergence to the stationary solutions of equation~(\ref{eq:Prob-NAND}). In the range $0<\epsilon<\epsilon^*$; for $\Delta=\epsilon^*-\epsilon$ we find
\begin{eqnarray}
\textrm{const}\times\exp\left[\frac{\ell}{2}\log\left(8\left(\frac{3\!-\!\sqrt{7}}{4}\!-\!\Delta\right)\left(\frac{3\!+\!\sqrt{7}}{4}\!+\!\Delta\right)\right)\right]\leq\vert P_-(\ell)-P_-(\infty)\vert\label{eq:ineq-NAND}.
\end{eqnarray}
Close to the critical noise level ($\Delta\rightarrow0$) the argument of the log function approaches unity and the number of layers needed to converge to the stationary solution (the point of intersection of all magnetization curves in Figure~\ref{fig:14})  diverges.
In the opposite limit of $\Delta\rightarrow\epsilon^*=\frac{3\!-\!\sqrt{7}}{4}$ the argument of the log function approaches zero and the convergence to the stationary states $m(\infty)=\pm1$ is very fast.

Finally, we note that the results of this section give us a positive answer to some of the conjectures put forward in~\cite{Evans:MTN}, in particular: a) The threshold $\epsilon^*=\frac{3\!-\!\sqrt{7}}{4}$ is valid for the random NAND formulae with completely reliable (hard) inputs. b)
The computation at $\epsilon=\epsilon^*$ is not possible, because equation (\ref{eq:Prob-NAND}) has only one fixed point which is both unstable and attractive.

\subsection{AND/OR gates\label{subsection:AND/OR}}

In this section we study random Boolean formulae constructed from a noisy AND and OR gates. The noiseless version of the AND/OR model defined on unbalanced trees was used in past to define the probability distribution on Boolean functions~\cite{Lefmann,Chauvin,Gardy}. The case of balanced trees was considered recently in~\cite{Fournier}.

 In the model of computation which we consider here the gate  $\alpha$ is sampled from the distribution $P(\alpha)=p\delta_{\alpha;\sgn[S_1+S_2+1]}+(1-p)\delta_{\alpha;\sgn[S_1+S_2-1]}$. Using this definition in the gate-averages of equations~(\ref{eq:m}) and (\ref{eq:Cl}) gives 
	\begin{eqnarray}
&&m(\ell+1)=\frac{1}{2}(1-2\epsilon)\left[2p-1+2m(\ell)-(2p-1)m^2(\ell)\right]\label{eq:m-AndOr}\\
&&C(\ell+1)=\frac{1}{4}(1-2\epsilon)\big[1\!-\!4 m(\ell)\!-\!4\hat m(\ell)\!+\!4C(\ell)\!+\!4p m(\ell)\!+\!4p\hat m(\ell)\label{eq:C-AndOr}\\
&&\!-\!4p m(\ell)C(\ell)\!-\!4p\hat m(\ell)C(\ell)\!-\!m^2(\ell)\!+\!4m(\ell)\hat m(\ell)\!+\!4m(\ell)C(\ell)\!-\!\hat{m}^2(\ell)\!+\!4\hat m(\ell)C(\ell)\!+\!C^2(\ell)\big]\nonumber
\end{eqnarray}
Equation~(\ref{eq:m-AndOr}) can be written in a more convenient form
\begin{eqnarray}
P_{-}(\ell+1)=\epsilon+2(1-2\epsilon)(1-p)P_{-}(\ell)+(1-2\epsilon)(2p-1)P_{-}^2(\ell)\label{eq:Prob-AndOR}
\end{eqnarray}
where $P_{-}(\ell)=(1-m(\ell))/2$ is the probability of output taking value of $-1$. For $\epsilon>0$  equation~(\ref{eq:Prob-AndOR}) has only one (stable) steady state solution $P_{-}(\infty)$. Thus there is no phase transition in this model for any noise value $\epsilon>0$ and the information about the input cannot be preserved for infinitely many layers.

The noiseless balanced AND/OR trees were studied in~\cite{Fournier}. Here we only show how to recover their results~\footnote{To establish the connection between~\cite{Fournier} and our work we use the mapping $S_i=1-2x_i$ from $x_i\in\{1,0\}$ to $S_i\in\{-1,1\}$} from the equations~(\ref{eq:m-AndOr}), (\ref{eq:C-AndOr}) and (\ref{eq:Prob-AndOR}).

Firstly, we note that by setting $\epsilon=0$ in  equation~(\ref{eq:Prob-AndOR}) one obtains the equation of Lemma 3.1 in~\cite{Fournier}. Equation~(\ref{eq:Prob-AndOR}) has two fixed points $P_{-}(\infty)\in\{1,0\}$($m(\infty)\in\{-1,1\}$) when $p\neq1/2$. The first point $P_{-}(\infty)=1$ is stable while the second point $P_{-}(\infty)=0$ is unstable when $p<1/2$, so the circuit computes the OR function of the variables belonging to the input set $S^I$. For  $p>1/2$ the point $P_{-}(\infty)=1$ is unstable and the point $P_{-}(\infty)=0$ is stable, so the circuit computes AND function.

Secondly, we set $p=1/2$ and allow for $m(0)\neq\hat m(0)$, i.e. we have two copies of the same circuit but with different inputs in the equations~(\ref{eq:m-AndOr}), (\ref{eq:C-AndOr}). For $p=1/2$ the magnetization in the circuit is conserved from layer to layer ($m(\ell)=m(0)$) and the overlap equation~(\ref{eq:C-AndOr}) reduces to
 \begin{eqnarray}
C(\ell+1)=\frac{1}{4}+\frac{1}{2}C(\ell)-\frac{1}{4}m^2(0)+\frac{1}{2}m(0)\hat m(0)-\frac{1}{4}\hat m^2(0)+\frac{1}{4}C^2(\ell).
\end{eqnarray}
Using the above equation to compute the joint probability $P^{\ell}(-1,1)=(1-m(0)+\hat m(0)-C(\ell))/4$ gives 
 \begin{eqnarray}
P^{\ell}(-1,1)=P^{\ell}(-1,1)\left[1-\frac{1}{2}m(0)+\frac{1}{2}\hat m(0)-P^{\ell}(-1,1)\right]~, \label{eq:Prob-11-AndOR}
\end{eqnarray}
which is the equation of Lemma 3.2 in~\cite{Fournier}. In general, for $\ell\rightarrow\infty$, we have $P^{\infty}(S,\hat S)=(1+Sm(0)+\hat S\hat m(0)+S\hat SC(\infty))/4$, where $C(\infty)=1-\vert\hat m(0)-m(0)\vert$. The analysis~\cite{Fournier} of equation~(\ref{eq:Prob-11-AndOR}) reveals that AND/OR-based formulae compute constant $\pm1$ functions when $S^I=\{S_1,\ldots,S_n,-S_1,\ldots,-S_n\}$ and linear threshold functions $\sgn[\sum_{j=1}^n S_j-n+2i]$, where $i\in\{1,\ldots,n\}$ and $S^I=\{S_1,\ldots,S_n\}$. Finally, the convergence to the functions computed in this model is mainly exponential in $\ell$ except in one special case of $p=1/2$ when it is logarithmic ($\sim\ell^{-1}$)~\cite{Fournier}.

\section{Discussion\label{section:Discussion}}

We have presented the theoretical framework that allows one to study the typical properties of noisy random Boolean formulae. A Boolean formula is a simple model of computation, which plays an important	role in many areas of the theoretical computer science (TCS). The tree-like structure of Boolean formulae allows for the  computation of exact noise thresholds by considering formulae with the worst possible topology. Another area of TCS where Boolean formulae played an important role is the generation of random Boolean functions. Here one usually uses a growth process to generate random formulae that induce a uniform probability distribution on the Boolean functions they compute.

Here, for the first time, random formulae generated by the growth process are used to study the typical properties of noisy formulae. The method used here relies on the layered variant of the Savick\'{y} formula-growth process. The layered framework allows us a direct mapping to the physical Ising spin system, which can be seen as a dynamical system where the time-steps correspond to circuit layers. This analogy allows one to use the generating functional analysis (GFA) method of statistical physics. The GFA method has an excellent record in the area of  disordered dynamical systems and is generally accepted to be exact.
	
Here, we use GFA to study the typical properties of noisy random Boolean formulae constructed from single gates or distributions of gates. All exact noise thresholds, which were derived  in TCS using rigorous methods, are recovered within our framework and identified with the corresponding macroscopic phase transitions. We attribute this to the exact correspondence of the mean-field equation~(\ref{eq:m}) to the single-gate probability of error (assuming independence of inputs) used to derive these noise thresholds in TCS. However, many of the properties of noisy random Boolean formulae studied here are inaccessible via the traditional analytic methods of TCS and Information Theory.

In the noiseless case ($\epsilon=0$), we have identified the Boolean functions generated by the layered growth process, but only when the input-set magnetization $m(0)=0$ is not a fixed point of the dynamics~(\ref{eq:m}). For inputs with $m(0)=0$ our results are consistent with the results of the Savick\'{y} growth process that can generate a Boolean function of arbitrary complexity. Furthermore, we have established that the functions generated in the layered growth process are sensitive to the input variables, an indication of their complexity. In order to find out exactly which functions are generated for the input set with $m(0)=0$, or in a more general setting with $\epsilon>0$, one needs a more sophisticated version of the mean-field theory presented here~\cite{BooleanF}.

For $\epsilon>0$ we have studied the evolution of the output magnetization and computation-error from layer to layer  and their dependence on the input-set magnetization and noise. We have identified a range of gate-noise parameter, and its dependence on the input-set magnetization $m(0)$, where by adding more layers to the circuit one can reduce the computation error. The speed of convergence to the equilibrium was studied both numerically (for $\epsilon=0$) and analytically (for $\epsilon>0$). For $\epsilon=0$ our numerical results are consistent with the rigorous bounds, but show that in a typical case a much tighter bound can be derived.

The standard noisy computation model was expanded to include "production errors" (uncorrelated hard noise). We have found that the effect of hard noise on the critical behavior of noisy circuit is to effectively reduce its critical noise threshold. We expect that the critical noise threshold (if it exists) in any noisy circuit will be affected in this way. Also, in our work a standard flip-noise is compared with  threshold noise. In particular, we have found that the MAJ-$k$ gate is more resilient to  threshold noise than to  flip-noise. We expect that any gate which can be represented as the linear threshold function is more robust against the threshold noise, at least for the distribution of threshold noise considered in this paper.

We believe that much can be learned about the typical properties of noisy computation via this approach, which complements the rigorous bounds derived in the TCS literature and provides insight that may help in the development of new rigorous techniques. 

\begin{acknowledgments} Support by the Leverhulme trust (grant F/00 250/H) is gratefully acknowledged. \end{acknowledgments}
\appendix

\section{Noise average\label{appendix:Process}}

In this appendix, we show for completeness how to derive the microscopic law (\ref{eq:micro}) from the  basic computation step in the von Neumann's model of noisy computation. The basic step in a noisy circuit is to compute the output of the ($\ell,i$)-th gate, given the input $\Spin_{i_1}^{\ell-1},\ldots,\Spin_{i_k}^{\ell-1}$, according to the stochastic rule
	\begin{eqnarray}
\Spin_{i}^\ell=\eta_{i}^\ell\alpha_i^\ell
(\Spin_{i_1}^{\ell-1},\ldots,\Spin_{i_k}^{\ell-1})\label{def:algorithm}
\end{eqnarray}
where $\eta_{i}^\ell$ is an independent random variable from the distribution  $\Prob(\eta)=\epsilon\delta_{\eta;-1}\!+\!(1\!-\!\epsilon)\delta_{\eta;1}$. Equation (\ref{def:algorithm}) gives rise to the conditional probability
\begin{eqnarray}
\Prob(\Spin_{i}^\ell\vert \Spin_{i_1}^{\ell-1},\ldots,\Spin_{i_k}^{\ell-1})=\left\langle\delta_{S_{i}^\ell;\,\eta\alpha_i^\ell (S_{i_1}^{\ell-1},\ldots,S_{i_k}^{\ell-1})}\right\rangle_{\eta}.\label{def:condProb}
\end{eqnarray}
Averaging out the noise variable  produces the equation
\begin{eqnarray}
\Prob(\Spin_{i}^\ell\vert \Spin_{i_1}^{\ell-1},\ldots,\Spin_{i_k}^{\ell-1})=\frac{1}{2}[1\!+\!(1\!-\!2\epsilon)S_{i}^\ell\alpha_i^\ell
(S_{i_1}^{\ell-1},\ldots,S_{i_k}^{\ell-1})]~, \label{eq:condProb}
\end{eqnarray}
which in turn leads to the equation (\ref{eq:micro}) if one uses the transformation of noise variables $1\!-\!2\epsilon=\tanh\beta$, where $\beta\in[0,\infty)$, and exploit the property $-\tanh(x)=\tanh(-x)$. 

\section{Disorder average\label{appendix:Averages}}

Here, we outline the calculation steps which lead to the saddle-point integral (\ref{eq:integral}). The starting point of this calculation is the generating functional
	\begin{eqnarray}
\Gamma[\vecPsi;\vecPsihat]&=&\sum_{\{\bf{S}^\ell,\hat{\bf{S}}^\ell\}}\Prob(\vecSpin^0,\vecSpinhat^0\vert \vecSpin^I)
\exp\left[\sum_{\ell,i=1}^{L,N} \left\{\beta \Spin_{i}^\ell H_i^{\ell-1}(\vecSpin^{\ell-1}) + \hat\beta \Spinhat_{i}^\ell\hat{H}_i^{\ell-1}(\vecSpinhat^{\ell-1})\right\}\right]\label{eq:Zav1}\\
&&\times\exp\left[-\sum_{\ell,i=1}^{L,N} \left\{
 \log2\cosh[\beta H_i^{\ell-1}(\vecSpin^{\ell-1})] +\log2\cosh[\hat\beta\hat{H}_i^{\ell-1}(\vecSpinhat^{\ell-1})]\right\}\right]\nonumber\\
&&\times\rme^{-\rmi\sum_{\ell,i}\{\psi_i^{\ell} S_{i}^{\ell}+\Psihat_i^{\ell} \hat{S}_{i}^{\ell}\}}\nonumber
\end{eqnarray}
where in the above we have defined the field terms
$H_i^{\ell-1}(\vecSpin^{\ell-1})=\sum_{j_1,\ldots,j_k}^N A_{j_1,\ldots,j_k}^{\ell,i}\alpha_i^\ell (\Spin_{j_1}^{\ell-1},\ldots,\Spin_{j_k}^{\ell-1})$ and $\hat{H}_i^{\ell-1}(\vecSpinhat^{\ell-1})=\sum_{j_1,\ldots,j_k}^N A_{j_1,\ldots,j_k}^{\ell,i}\alpha_i^\ell (\Spinhat_{j_1}^{\ell-1},\ldots,\Spinhat_{j_k}^{\ell-1})$. Enforcing the definitions of fields in the equation~(\ref{eq:Zav1}) via the integral representations of unity
\begin{eqnarray}
\prod_{i=1}^{N}\prod_{\ell=0}^{L-1}\left\{\int\frac{\mathrm d H_i^\ell\mathrm d x_i^\ell}{2\pi}\rme^{\rmi x_i^\ell[H_i^\ell\!-\!H_i^\ell(\bf{S}^{\ell})]}\right\}\!=\!
\prod_{i=1}^{N}\prod_{\ell=0}^{L-1}\left\{\int\frac{\mathrm d \hat{H}_i^\ell\mathrm d \hat{x}_i^\ell}{2\pi}\rme^{\rmi \hat x_i^\ell[\hat H_i^\ell\!-\!\hat H_i^\ell(\hat{\bf{S}}^{\ell})]}\right\}\!=\!1\label{def:unity}
\end{eqnarray}
leads to
\begin{eqnarray}
\Gamma[\vecPsi;\vecPsihat]&=&\prod_{i=1}^{N}\prod_{\ell=0}^{L-1}\left\{\int\frac{\mathrm d H_i^\ell\mathrm d x_i^\ell\mathrm d \hat{H}_i^\ell\mathrm d \hat{x}_i^\ell}{(2\pi)^2}\right\}
\sum_{\{\bf{S}^\ell,\hat{\bf{S}}^\ell\}}\Prob(\vecSpin^0,\vecSpinhat^0\vert \vecSpin^I)\,\rme^{-\rmi\sum_{\ell,i}\{\psi_i^{\ell} S_{i}^{\ell}+\Psihat_i^{\ell} \hat{S}_{i}^{\ell}\}}\label{eq:Zav2}\\
&&\times\rme^{\sum_{i=1}^{N} \sum_{\ell=0}^{L-1}\left\{\beta S_{i}^{\ell+1} H_i^\ell + \hat\beta \hat{S}_{i}^{\ell+1}\hat{H}_i^\ell- \log2\cosh[\beta H_i^\ell] -\log2\cosh[\hat\beta\hat{H}_i^\ell]+ \rmi x_{i}^\ell H_i^\ell + \rmi\hat{x}_{i}^\ell\hat{H}_i^\ell\right\}    }\nonumber\\
&&\times\prod_{\ell,i=1}^{L,N}\prod_{j_1,\ldots,j_k}^{N}\rme^{-\rmi A_{j_1,\ldots,j_k}^{\ell,i} \{ x_{i}^{\ell-1}\alpha_i^\ell (S_{j_1}^{\ell-1},\ldots,S_{j_k}^{\ell-1}) + \hat{x}_{i}^{\ell-1} \alpha_i^\ell (\hat{S}_{j_1}^{\ell-1},\ldots,\hat{S}_{j_k}^{\ell-1})  \}}\label{eq:disorder}
\end{eqnarray}
For now we concentrate only on the last line of the equation~(\ref{eq:Zav2}) that depends on the connectivity~(\ref{def:connect-disorder}) and gate disorder~(\ref{def:gate-disorder}). We average over the disorder in~(\ref{eq:disorder}) as follows
\begin{eqnarray}
&&Z_A\!\sum_{\{A_{i_1,\ldots,i_k}^{\ell,i}\}}\Prob(\{A_{i_1,\ldots,i_k}^{\ell,i}\})\sum_{\{\alpha_{i}^{\ell}\}}\Prob(\{\alpha_{i}^{\ell}\})\label{eq:result}\\
&&\times\prod_{\ell,i=1}^{L,N}\prod_{j_1,\ldots,j_k}^{N}\rme^{-\rmi A_{j_1,\ldots,j_k}^{\ell,i} \{ x_{i}^\ell\alpha_i^\ell (S_{j_1}^{\ell-1},\ldots,S_{j_k}^{\ell-1}) + \hat{x}_{i}^\ell \alpha_i^\ell (\hat{S}_{j_1}^{\ell-1},\ldots,\hat{S}_{j_k}^{\ell-1})  \}}\nonumber\\
&&=\sum_{\{A_{i_1,\ldots,i_k}^{\ell,i}\}}\prod_{\ell,i=1}^{L,N}\Big\{
\int_{-\pi}^{\pi}\frac{\mathrm d \omega_i^\ell}{2\pi}\rme^{\rmi\omega_i^\ell}\prod_{i_1,\ldots,i_k}^{N}\left[\frac{1}{N^k}\delta_{A_{i_1,\ldots,i_k}^{\ell,i};1}+(1-\frac{1}{N^k})\delta_{A_{i_1,\ldots,i_k}^{\ell,i};0}\right] \nonumber\\
&&\times\sum_{\alpha_{i}^{\ell}}\Prob(\alpha_{i}^{\ell})\,\rme^{-\rmi A_{i_1,\ldots,i_k}^{\ell,i} \{ x_{i}^{\ell-1}\alpha_i^\ell (S_{i_1}^{\ell-1},\ldots,S_{i_k}^{\ell-1}) + \hat{x}_{i}^{\ell-1} \alpha_i^\ell (\hat{S}_{i_1}^{\ell-1},\ldots,\hat{S}_{i_k}^{\ell-1})  +\omega_i^\ell\}}\Big\}\nonumber\\
&&=\prod_{\ell,i=1}^{L,N}
\int_{-\pi}^{\pi}\frac{\mathrm d \omega_i^\ell}{2\pi}\rme^{\rmi\omega_i^\ell}\exp\left[\frac{1}{N^k}\sum_{i_1,\ldots,i_k}^{N} \left\langle\rme^{-\rmi  \{ x_{i}^{\ell-1}\alpha (S_{i_1}^{\ell-1},\ldots,S_{i_k}^{\ell-1}) + \hat{x}_{i}^{\ell-1} \alpha (\hat{S}_{i_1}^{\ell-1},\ldots,\hat{S}_{i_k}^{\ell-1})  +\omega_i^\ell\}}\!-\!1\right\rangle_{\alpha} \!+\!O(N^{-k})\right]\nonumber
\end{eqnarray}
In the first line of the calculation we have used the integral representation of Kronecker delta function $\delta_{n;m}=\int_{-\pi}^{\pi}\frac{\mathrm d \omega}{2\pi}\rme^{\rmi\omega(n-m)}$ and in the last line the exponential form is valid for large $N$. By setting all the $x_{i}^\ell$ and $\hat{x}_{i}^\ell$ variables to $0$ in the equation (\ref{eq:result}), we find that the normalization constant $Z_A$ contributes the factor $\rme^{NL}$ to the generating functional (\ref{eq:Zav1}) when $N$ is large. Using the result of disorder-average (\ref{eq:result}) in the equation (\ref{eq:Zav1}) we obtain the disorder-averaged generating functional
\begin{eqnarray}
\overline{\Gamma[\vecPsi;\vecPsihat]}&=&\prod_{i=1}^{N}\prod_{\ell=0}^{L-1}\left\{\int\frac{\mathrm d H_i^\ell\mathrm d x_i^\ell\mathrm d \hat{H}_i^\ell\mathrm d \hat{x}_i^\ell}{(2\pi)^2}\int_{-\pi}^{\pi}\frac{\mathrm d \omega_i^{\ell+1}}{2\pi}\rme^{\rmi\omega_i^{\ell+1}}\right\}\label{eq:Zav3}\\
&&\times\sum_{\{\bf{S}^\ell,\hat{\bf S}^\ell\}}\Prob(\vecSpin^0,\vecSpinhat^0\vert \vecSpin^I)\,\rme^{-\rmi\sum_{\ell,i}\{\psi_i^{\ell} S_{i}^{\ell}+\Psihat_i^{\ell} \hat{S}_{i}^{\ell}\}+NL}\nonumber\\
&&\times\rme^{\sum_{i=1}^{N} \sum_{\ell=0}^{L-1} \left\{\beta S_{i}^{\ell-1} H_i^\ell + \hat\beta \hat{S}_{i}^{\ell-1}\hat{H}_i^\ell- \log2\cosh[\beta H_i^\ell] -\log2\cosh[\hat\beta\hat{H}_i^\ell]+ \rmi x_{i}^\ell H_i^\ell + \rmi\hat{x}_{i}^\ell\hat{H}_i^\ell\right\}    }\nonumber\\
&&\times\prod_{\ell=1}^{L}
\exp\left[\frac{1}{N^k}\sum_{i,i_1,\ldots,i_k}^{N} \left\langle\rme^{-\rmi \{ x_{i}^{\ell-1}\alpha (S_{i_1}^{\ell-1},\ldots,S_{i_k}^{\ell-1}) + \hat{x}_{i}^{\ell-1} \alpha (\hat{S}_{i_1}^{\ell-1},\ldots,\hat{S}_{i_k}^{\ell-1})  +\omega_i^\ell\}}\!-\!1\right\rangle_{\alpha} \!+\!O(N^{-k+1})\right]\nonumber
\end{eqnarray}
In order to achieve factorization over sites in the equation we isolate the densities
\begin{eqnarray}
 &&\Prob^\ell(S,\hat S)=\frac{1}{N}\sum_{i=1}^N\delta_{S;S_{i}^{\ell}}\delta_{\hat S;\hat{S}_{i}^{\ell}}\\
&&\Omega^\ell(x,\hat{x},\omega)=\frac{1}{N}\sum_{i=1}^N\delta(x\!-\!x_i^\ell)\delta(\hat{x}\!-\!\hat{x}_i^\ell)\delta(\omega\!-\!\omega_i^{\ell+1})
\end{eqnarray}
via the respective integro-functional representations of unity
\begin{eqnarray}
&&\int\{\mathrm d P^\ell \mathrm d\hat{P}^\ell\}\rme^{\rmi N\!\sum_{S,\hat S}\hat{P}^\ell(S,\hat S)[P^\ell(S,\hat S)-\frac{1}{N}\sum_{i=1}^N\delta_{S;S_{i}^{\ell}}\delta_{\hat S;\hat{S}_{i}^{\ell}}]}=1\\
&&\int\{\mathrm d\Omega^\ell \mathrm d\hat{\Omega}^\ell\}\rme^{\rmi N\! \int\mathrm d x\mathrm d \hat{x}\mathrm d \omega\hat{\Omega}^\ell(x,\hat{x},\omega)[\Omega^\ell(x,\hat{x},\omega)\!-\!\frac{1}{N}\sum_{i=1}^N\delta(x\!-\!x_i^\ell)\delta(\hat{x}\!-\!\hat{x}_i^\ell)\delta(\omega\!-\!\omega_i^{\ell+1})]}=1\nonumber
\end{eqnarray}
which leads to
\begin{eqnarray}
&&\overline{\Gamma[\vecPsi;\vecPsihat]}\label{eq:Zav4}\\
&&=\int\{\mathrm d\vecProb \mathrm d\vecProbhat\mathrm d\vecOmega \mathrm d\vecOmegahat\}\exp\left[N\sum_{\ell=0}^{L-1} \left\{ \rmi\!\sum_{S,\hat S}\Probhat^{\ell}(S,\hat S)\Prob^{\ell}(S,\hat S)+    \rmi\! \int\mathrm d x\mathrm d \hat{x}\mathrm d \omega\hat{\Omega}^\ell(x,\hat{x},\omega)\Omega^\ell(x,\hat{x},\omega) \right\}  \right]\nonumber\\
&&\times\exp\left[N\sum_{\ell=0}^{L-1}\sum_{\{S_j,\hat{S}_j\}}\prod_{j=1}^k\left\{P^{\ell}(S_j,\hat{S}_j)\right\}
\int\mathrm d x\mathrm d \hat{x}\mathrm d \omega\Omega^\ell(x,\hat{x},\omega)
\left\langle\rme^{-\rmi \{ x\alpha (S_{1},\ldots,S_{k}) \!+\! \hat{x} \alpha (\hat{S}_{1},\ldots,\hat{S}_{k})\!+\!\omega\}}\right\rangle_{\alpha} \right]\nonumber\\
&&\times\prod_{i=1}^N\Bigg[\sum_{S_i^0,\hat{S}_i^0}\delta_{S_i^0;S_{n_i}^I}\delta_{S_i^0;\hat{S}_{i}^0}\prod_{\ell=0}^{L-1}\Bigg\{\sum_{S_i^\ell,\hat{S}_i^\ell}\int\frac{\mathrm d H_i^\ell\mathrm d x_i^\ell\mathrm d \hat{H}_i^\ell\mathrm d \hat{x}_i^\ell}{(2\pi)^2}\int_{-\pi}^{\pi}\frac{\mathrm d \omega_i^{\ell+1}}{2\pi}\rme^{\rmi\omega_i^{\ell+1}}\nonumber\\
&&\times\rme^{\beta S_{i}^{\ell+1} H_i^\ell + \hat\beta \hat{S}_{i}^{\ell+1}\hat{H}_i^\ell- \log2\cosh[\beta H_i^\ell] -\log2\cosh[\hat\beta\hat{H}_i^\ell]+ \rmi x_{i}^\ell H_i^\ell + \rmi\hat{x}_{i}^\ell\hat{H}_i^\ell -\rmi \hat{P}^{\ell}(S_{i}^{\ell},\hat{S}_{i}^{\ell})-\rmi\hat{\Omega}^\ell(x_i^\ell,\hat{x}_i^\ell,\omega_i^{\ell+1})   }\Bigg\}\nonumber\\
&&\times\rme^{-\rmi\sum_{\ell=0}^L\{\psi_i^{\ell} S_{i}^{\ell}+\Psihat_i^{\ell} \hat{S}_{i}^{\ell}\}}\Bigg]\rme^{O(L N^{-k+1})}\nonumber
\end{eqnarray}
The site-dependent part of the above equation can be written in the form
\begin{eqnarray}
\exp\!\left[\!\sum_n\frac{1}{N}\sum_{i=1}^N\delta_{n;n_i}\!\log\!\int\!\{\mathrm d\vecField_i\mathrm d\vecconjField_i\mathrm d\hat\vecField_i\mathrm d\hat\vecconjField_i\}\!\int\!\mathrm D\vecomega_i\!\sum_{\textbf{S}_i,\hat{\textbf{S}}_i}\! M_{n}[\vecField_i,\vecconjField_i;\hat\vecField_i,\hat\vecconjField_i;\vecomega_i;\vecSpin_i,\vecSpinhat_i]\right]\label{eq:site}
\end{eqnarray}
where we have defined the effective single-site measure
\begin{eqnarray}
&&M_{n_i}[\vecField_i,\vecconjField_i;\hat\vecField_i,\hat\vecconjField_i;\vecomega_i;\vecSpin_i,\vecSpinhat_i]\\
&&=\delta_{S_i^0;S_{n_i}^I}\delta_{S_i^0;\hat{S}_{i}^0}\,\rme^{-\rmi\sum_{\ell=0}^L\{\psi_i^{\ell} S_{i}^{\ell}+\Psihat_i^{\ell} \hat{S}_{i}^{\ell}\}}\nonumber\\
&&\times\prod_{\ell=0}^{L-1}\rme^{\beta S_{i}^{\ell+1} H_i^\ell + \hat\beta \hat{S}_{i}^{\ell+1}\hat{H}_i^\ell- \log2\cosh[\beta H_i^\ell] -\log2\cosh[\hat\beta\hat{H}_i^\ell]+ \rmi x_{i}^\ell H_i^\ell + \rmi\hat{x}_{i}^\ell\hat{H}_i^\ell -\rmi \hat{P}^{\ell}(S_{i}^{\ell},\hat{S}_{i}^{\ell})-\rmi\hat{\Omega}^\ell(x_i^\ell,\hat{x}_i^\ell,\omega_i^{\ell+1})   }\nonumber
\end{eqnarray}
and we use the definition $\int\mathrm D\vecomega_i=\prod_{\ell=1}^{L}\int_{-\pi}^{\pi}\frac{\mathrm d \omega_i^\ell}{2\pi}\rme^{\rmi\omega_i^\ell}$. Using the definition (\ref{eq:site}) in the disorder-averaged generating functional (\ref{eq:Zav4}) with all the generating fields $\{\psi_i^{\ell},\Psihat_i^{\ell}\}$ being set to $0$ and assuming that the law of large numbers for the random index-variables $\{n_i\}$ holds, i.e. $\lim_{N\rightarrow\infty}\frac{1}{N}\sum_{i=1}^N\delta_{n;n_i}\rightarrow\Prob(n)$, we arrive at the result of equation~(\ref{eq:integral}).

\section{Simplification of the saddle-point problem\label{appendix:SPproblem}}

In this appendix, we show how to solve the saddle-point equations~(\ref{eq:SP1})-(\ref{eq:SP4}). First, we use the saddle-point equation (\ref{eq:SP4}) to eliminate the conjugate order parameter $\hat{\Omega}^\ell$ from the effective measure (\ref{eq:M}), giving 
\begin{eqnarray}
\M_n[\ldots]&=&\delta_{S^0;S^I_n}\delta_{\hat{S}^0;S^0}\prod_{\ell=0}^{L-1}
\frac{\rme^{\beta S^{\ell+1}\Field^{\ell}+\hat\beta\hat{S}^{\ell+1}\Fieldhat^{\ell}}}{2\cosh(\beta\Field^\ell)2\cosh(\hat\beta\Fieldhat^\ell)}
\,\rme^{\rmi\conjField^\ell\Field^\ell+\rmi\conjFieldhat^\ell\Fieldhat^\ell+\rmi\omega^{\ell+1}} \\
&&\times\exp\left[\sum_{\{S_j,\hat{S}_j\}}\prod_{j=1}^k\left[\Prob^\ell( S_j,\hat{S}_j)\right]
\left\langle\rme^{-\rmi\{x^\ell\alpha (\{S_j\})+\hat{x}^\ell\alpha (\{\hat{S}_j\})+\omega^{\ell+1}\}}\right\rangle_{\alpha}
\right]\nonumber\\
&&\times\rme^{-\rmi\Probhat^\ell( S^\ell,\hat{S}^\ell)} ~. \nonumber
\end{eqnarray}
Second, using the above result we compute the Fourier transform
\begin{eqnarray}
F_\gamma^{\ell^{\prime}}[y,z]=\int\{\rmd\vecField\rmd\vecconjField\rmd\hat\vecField\rmd\hat\vecconjField \mathrm D\vecomega\}\sum_{\vecSpin,\vecSpinhat}\M_n[\vecField,\vecconjField;\hat\vecField,\hat\vecconjField;\vecomega;\vecSpin,\vecSpinhat]\rme^{-\rmi\gamma\{x^{\ell^\prime} y+\hat{x}^{\ell^\prime} z+\omega^{\ell^\prime+1}\}}\label{def:Fourier}
\end{eqnarray}
where $\gamma\in\{0,1\}$. For $\gamma=0$ we obtain
\begin{eqnarray}
F_0^{\ell^{\prime}}[y,z]=\sum_{\{S^\ell,\hat{S}^\ell\}}\delta_{S^0;S^I_n}\delta_{\hat{S}^0;S^0}\prod_{\ell=0}^{L-1}W[S^{\ell+1};\hat{S}^{\ell+1}]N[S^\ell;\hat{S}^\ell]\label{def:F0}
\end{eqnarray}
and for $\gamma=1$ we have
\begin{eqnarray}
F_1^{\ell^{\prime}}[y,z]=\sum_{\{S^\ell,\hat{S}^\ell\}}\delta_{S^0;S^I_n}\delta_{\hat{S}^0;S^0}\tilde W[S^{\ell^{\prime}+1};\hat{S}^{\ell^{\prime}+1}]N[S^{\ell^\prime};\hat{S}^{\ell^\prime}]\prod_{\ell\neq\ell^\prime}^{L-1}W[S^{\ell+1};\hat{S}^{\ell+1}]N[S^\ell;\hat{S}^\ell]\label{def:F1}
\end{eqnarray}
where
\begin{eqnarray}
&&W[S^{\ell+1};\hat{S}^{\ell+1}]=\sum_{\{S_j,\hat{S}_j\}}\prod_{j=1}^k\left[\Prob^\ell( S_j,\hat{S}_j)\right] \left\langle
\frac{\rme^{\beta S^{\ell+1}\alpha (\{S_j\})+\hat\beta\hat{S}^{\ell+1}\alpha (\{\hat{S}_j\})}}{2\cosh(\beta \alpha (\{S_j\}))2\cosh(\hat\beta \alpha (\{\hat{S}_j\}))}
\right\rangle_{\alpha}\label{def:W1}\\
&&\tilde W[S^{\ell+1};\hat{S}^{\ell+1}]=\frac{\rme^{\beta S^{\ell+1}y+\hat\beta\hat{S}^{\ell+1}z}}{2\cosh(\beta y)2\cosh(\hat\beta z)}\label{def:W2}\\
&&N[S^\ell;\hat{S}^\ell]=\rme^{-\rmi\Probhat^\ell( S^\ell,\hat{S}^\ell)}\label{def:N}.
\end{eqnarray}
%
Next we notice that $\int\rmd x \rmd \hat x \rmd\omega\Omega^{\ell^\prime}(x, \hat x,\omega)\rme^{-\rmi\{x y+\hat x z+\omega\}}=\frac{F_1^{\ell^{\prime}}[y,z]}{F_0^{\ell^{\prime}}[y,z]}$. Using the fact that (\ref{def:W1}) and (\ref{def:W2}) are both probability distributions the computation for $\ell^\prime=L-1$ gives $\frac{F_1^{\ell^{\prime}}[y,z]}{F_0^{\ell^{\prime}}[y,z]}=1$. Plugging in this result into the saddle-point equation (\ref{eq:SP2}) gives us $\Probhat^{L-1}( S,\hat{S})=\rmi k$ implying that $N[S^{L-1};\hat{S}^{L-1}]=\rme^{k}$. The latter is used to show that for $\ell^\prime=L-2$ gives $\frac{F_1^{\ell^{\prime}}[y,z]}{F_0^{\ell^{\prime}}[y,z]}=1$ and so on until we conclude that $\Probhat^{\ell}( S,\hat{S})=\rmi k$ for all $\ell$.

\end{document}